\documentstyle[aps,prl,multicol,epsf]{revtex}
\begin{document}
\def\Re{\, {\cal R}\mkern-3.1mu e\,}
\def\B.#1{{\bbox{#1}}}
\def\C.#1{{\cal{#1}}}
\def\BC.#1{\bbox{\cal{#1}}}
\def\sub.#1,#2 {_{\hbox{\tiny#1,{\scriptsize#2}}}}
\def\Z.#1 {{\zeta^{^{\rm   K41}}_{#1}}}
\def\sb.#1 {\lower0.40ex\hbox{\tiny#1}}
\def\sb.#1 {_{\hbox{\tiny#1}}}
\def\sp.#1 {^{\hbox{\tiny#1}}}
\def\BE.#1 {\begin{equation}\label{#1}}
\def\BEA.#1 {\begin{eqnarray}\label{#1}}
\def\EE{\end{equation}}
\def\EEA{\end{eqnarray}}
\def\la {\langle}
\def\ra {\rangle}
\def\nn {\nonumber}
\def\BR{  \nonumber \\ }
\def\br{  \\ \nonumber }
\def \RRA{\big\rangle\!\!\big\rangle}
\def \LLA{\big\langle\!\!\big\langle}
\def\Fbox#1{\vskip1ex\hbox to 8.5cm{\hfil\fboxsep0.3cm\fbox{%
                    \parbox{8.0cm}{#1}}\hfil}\vskip1ex}
\def\QM{\fboxrule0.2ex\fbox{\large\bf ?}}
\def\EP{\fboxrule0.2ex\fbox{\large\bf !}}
\def\K41{\fboxrule0.2ex\fbox{\large\bf K41}}
\def\tr {(t,\B.r)}
\def\tR {(t,\B.R)}
\def\tx {(t,\B.x)}
\def\hf{\case{1}{2}}
\def\e{\epsilon}
\def\a{\alpha}
\def\o{\omega}
\def\O{\Omega}
\def\b{\beta}
\def\d{\delta}
\def\g{\gamma}
\def\k{\kappa}
\def\L{\Lambda}
\def\l{\lambda}
\def\t{\tau}
\def\z{\zeta}
\def\P{\Phi}
\def\Int {\int \limits_{-\infty}^\infty}

\title{{\rm PRE, submitted\hfill {\sl  \today}} \\
Analytic Calculation of the  Anomalous Exponents in Turbulence:\\
Using the Fusion Rules to Flush Out a Small Parameter}
  \author {Victor S. L'vov and Itamar  Procaccia}
  \address{Department of~~Chemical Physics,
 The Weizmann Institute of Science,
  Rehovot 76100, Israel}
 \maketitle

\begin{abstract}
The main difficulty of statistical theories of fluid turbulence is the
lack of an obvious small parameter. In this paper we show that the
formerly established fusion rules can be employed to develop a theory
in which Kolmogorov's statistics of 1941 acts as the zero order, or
background statistics, and the anomalous corrections to the K41
scaling exponents $\zeta_n$ of the $n$th order structure functions can
be computed analytically.  The crux of the method consists of
renormalizing a 4-point interaction amplitude on the basis of the
fusion rules. The novelty is that this amplitude includes a small
dimensionless parameter, which is shown to be of the order of the
anomaly of $\zeta_2$, $\delta_2= \zeta_2-2/3\approx 0.03$. Higher
order interaction amplitudes are shown to be even smaller. The
corrections to K41 to $O(\delta_2)$ result from standard
logarithmically divergent ladder-diagrams in which the 4-point
interaction acts as a ``rung". The theory allows a calculation of the
anomalous exponents $\zeta_n$ in powers of the small parameter
$\delta_2$. The $n$-dependence of the scaling exponents $\zeta_n$
stems from pure combinatorics of the ladder diagrams. In this paper we
calculate the exponents $\zeta_n$ up to $O(\delta_2^3)$. Previously
derived bridge relations allow a calculation of the anomalous
exponents of correlations of the dissipation field and of dynamical
correlations in terms of the same parameter $\delta_2$. The actual
evaluation of the small parameter $\delta_2$ from first principles
requires additional developments that are outside the scope of this
paper.
\end{abstract}
\begin{multicols}{2}
\section{Introduction}
\label{s:intro}
The aim of this paper is to build on previous work to achieve a
controlled evaluation of the anomalous exponents that characterize
various correlation and structure function in Navier-Stokes
turbulence, and in particular the exponents $\zeta_n$ that
characterize $n$th order structure functions.  The main result of this
paper is that given a single experimental input (for example the value
of the anomalous exponent of the second order structure function), the
$n$ dependence of all the other exponents that were reliably measured
in experiments and simulations can be calculated analytically.

Decades of experimental and theoretical attention (see for example
\cite{41Kol,MY-2,84AGHA,93SK,93Ben,Fri}) have been devoted to two
 types of simultaneous correlation functions; the first type includes
 the structure functions of velocity differences,
\begin{equation}
S_n(\B.R)= \left\langle | \B.u (\B.r+\B.R)- \B.u(\B.r)
|^n \right \rangle \,, \label{Sns}
\label{du1}
\end{equation}
where $\left<\dots\right>$ stands for a suitably defined ensemble
average.  A second type of correlations include gradients of the
velocity field.  An important example is the rate $\epsilon({\B.
r},t)$ at which energy is dissipated into heat due to viscous
damping. This rate is roughly $ \nu|\nabla {\B.u}({\B.r},t)|^2$.  An
often-studied simultaneous correlation function of
$\hat\epsilon({\B.r},t) =
\epsilon ({\B.r},t) - \bar\epsilon$ is
\begin{equation}
K_{\epsilon\epsilon}(\B.R) = \left< \hat\epsilon ({\B.r}+{\B.R})
\hat\epsilon ({\B.r}) \right> \ . \label{Kee}
\end{equation}
It has been hypothesized by Kolmogorov in 1941 (K41) and 1962 (K62)
that statistical objects of this type exhibit power law dependence on
$R$ \cite{41Kol,62Kol}:
\begin{equation} \label{defzn}
S_n(R) \propto R^{\zeta_n} \ ,\quad
K_{\epsilon\epsilon}(R) \propto R^{-\mu} \ .
\end{equation}
In addition, the K41 theory predicted the values of $\zeta_n$ to be
$n/3$. Experimental measurements and computer simulations show that in
some aspects K41 was remarkably close to the truth.  The major aspect
of its predictions, that the statistical quantities depend on the
length scale $R$ as power laws, is corroborated by experiments. On the
other hand, the predicted exponents seem not to be exactly
realized. The numerical values of $\zeta_n$ deviate progressively from
$n/3$ when $n$ increases \cite{84AGHA,93Ben}.  K62 tried to improve on
this prediction by taking into account the fluctuations in the rate of
energy dissipation.  On the basis of a phenomenological model,
assuming the distributions function of energy dissipation to be
lognormal, K62 reached the predictions
\begin{equation}
\zeta_n=\frac{n}{3} -\frac{\mu n(n-3)}{18} \ . \label{K62}
\end{equation}
Besides the fact that these predictions did not follow from fluid
mechanical considerations, it was pointed out \cite{Fri} that they are
violating basic inequalities that do not allow the exponents $\zeta_n$
to decrease, something that always happens with (\ref{K62}) with $n$
large enough. The quest for computing the scaling exponents from the
equations of fluid mechanics was long, arduous, and on the whole
pretty unsuccessful.

\begin{figure}
\epsfxsize=8.6truecm
\epsfbox{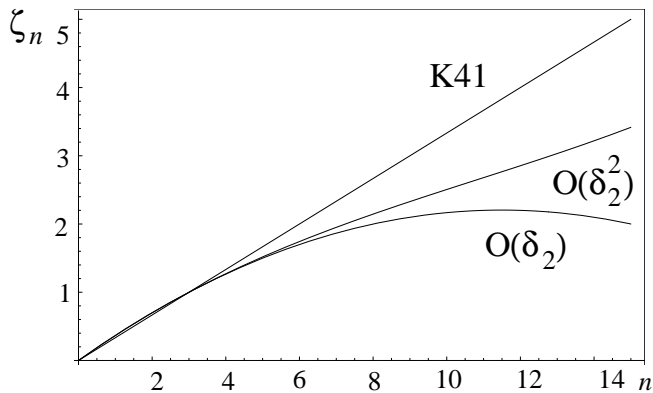}
\narrowtext
\caption{The scaling exponents $\zeta_n$ as a function of $n$.
 The calculation is organized by the small parameter
 $\delta_2=\zeta_2-2/3\approx 0.03$.  Shown is the K41 prediction
 which is zero-order in $\delta_2$, together with our results to first
 and second order in $\delta_2$. To first order the results are the
 same as the phenomenological prediction of K62, and to 2nd order it
 is Eq.~(5) with $b_2= -0.55$ according to Eq.~(72).}
\label{fig-5}
\end{figure}

In this paper we present an approach that is based on our own previous
findings which culminates in the analytic calculation of exponents
like $\zeta_n$ and $\mu$. At present the calculation is not completely
from first principle. We need the input of {\em one} number from
experiment, say $\delta_2\equiv \zeta_2-2/3$. Given this number we can
calculate all the other exponents systematically with $\delta_2$ being
a small parameter that organizes our calculations. To first order in
$\delta_2$ we recapture (\ref{K62}). We will show that the result to
$O(\delta_2)$ is {\em universal}, independent of the details of the
calculations performed below.  To second order we find the
results
\begin{eqnarray}
\zeta_n&=&\frac{n}{3} -\frac{n(n-3)}{2}\delta_2[1
+2\delta_2(n-2)b_2]+O(\delta_2^3) \ ,
\label{finalzn1}\\
\mu&=&9\delta_2(1+8b_2\delta_2)+O(\delta_2^3)\ .
\end{eqnarray}
The curves of $\zeta_n$ vs. $n$ are shown in
Fig. \ref{fig-5}. We show the K41 prediction, the result of our
calculation to 1-loop order, and the 2-loop result that is presented
in Eq.~(\ref{finalzn1}). While the {\em form} of these results is
universal, the numerical value of the dimensionless parameter $b_2$
depends on the details of the calculations; we find that $b_2$ is
always negative and of the order of unity. Note that the 2-loop
results correct the unwanted down curving of the 1-loop calculation
(which is the same malaise as in K62).
In thinking about the strategy for this work we were led by some
insights that developed in the context of understanding how to compute
the scaling exponents of the Kraichnan model of passive scalar
advection \cite{68Kra,94Kra}. In that model a scalar field $T(\B.r,t)$
is advected by a Gaussian velocity field $\B.u(\B.r,t)$ which is
delta-correlated in time but which has a scaling exponent
$\zeta_2=\epsilon$. For $\epsilon=0$ the advected scalar has trivial
statistics, and for $\epsilon$ small the model has a natural small
parameter. It turned out that the calculation of the exponents can
proceed along two ways. The first, which is non-perturbative, was
pioneered in \cite{95GK}. It considers the differential equations that
the $n$th order correlation functions satisfy, and identifies the
anomalous scaling solutions as the zero modes of these differential
equations. The calculation of the exponents then depends on the
calculation of the zero modes themselves, a task that is not at all
easy, and therefore such calculations were never done for any order
but $O(\epsilon)$. In this method the renormalization scale is the
outer scale $L$, not the inner scale $\eta$, and the dimensionless
ratio of scales that carries the anomalous part of the exponents of
the structure functions is $L/R$ where $R$ is defined in
Eq.~(\ref{du1}).  A second method that was discussed in detail in
\cite{00ALPP} considers instead of the correlation functions the
averages of higher moments of $\nabla T$, {\em i.e.} $\langle |\nabla
T|^n\rangle$. These quantities diverge as powers of $L/\eta$, and the
exponents of this power are the same as the anomalous part of the
exponent of the $n$th order structure function.  The great advantage
of the second method is that one can write a perturbative theory in
$\epsilon$ for the scaling exponents themselves, {\em without any need
to compute the zero modes or any other functions of many variables}.
Thus the second method allows easily computations to $O(\epsilon^2)$
\cite{SP,epscube} and with some more efforts to higher orders.
 The insight gained is that one needs to focus on a quantity that
 offers the most feasible calculation, exposing the anomalous part of
 the exponents as it appears with a dimensionless ratio of length
 scales.

In Navier Stokes turbulence the situation is similar. On the one hand
we have a non-perturbative theory which in this case is the infinite
hierarchy formed by the equations of motion of the correlation
function \cite{98BLPP}. We can use this hierarchy to demonstrate that
anomalous solutions exist, but the computation of the scaling
exponents requires a calculation of the correlation functions
themselves. This is a very difficult task that up to now was not
accomplished in a satisfactory manner. The other approach will be
described in this paper. It will be a perturbative theory for the
scaling exponents themselves, not requiring the computation of the
correlation function along the way. Similarly to the second method in
the Kraichnan problem it will be based on considering limits of
correlation functions when $p$ coordinates fuse. In that limit we
create, even when all the distances are in the inertial range, a ratio
of large and small lengthscales that appears raised to the power of
the anomalous exponent $\zeta_n$.

The two previous findings that influence crucially the present
formulation are the mechanism for anomalous scaling that was announced
in Ref.\cite{LP-2}, and the fusion rules that were discussed in
Ref.\cite{96LP-1}. In short, Ref. \cite{LP-2} exposed the ladder
diagrams which appear in the theory of turbulence. These diagrams
contain logarithmic divergences that are summable to power laws with
anomalous exponents. These ladder diagrams contain ``rungs" of the
ladder, that are actually vertices with four, six and more ``legs", 
representing 4-point, 6-point and higher order interaction amplitudes. In
Ref.\cite{LP-2} these objects were represented in terms of infinite
series of diagrams that could not be resummed analytically. This is
where the fusion rules are now called to save the day. The fusion
rules determine the asymptotic properties of $n$-point correlation
functions when subgroups of $p$ coordinates coalesce together. As such
the fusion rules are non-perturbative, and are believed to be
exact. We use the fusion rules to determine the asymptotic properties
of the rungs. This is done such that a {\em calculation} of $\zeta_2$
from the theory will agree with the experimental value of
$\zeta_2$. We then show that the knowledge of the asymptotics suffices
for constructing a calculation of all the other scaling exponents, and
in particular of $\zeta_n$ for $n>2$ and of $\mu$. The crux is that in
the process of determining the analytic form of the rungs we discover
that their amplitude contains powers of a dimensionless small
parameter $\delta_2=\zeta_2-2/3\approx
0.03$. Using this small parameter in the renormalized 4-point
interaction allows us to develop a systematic expansion in orders of
$\delta_2$. At the end of this paper we sketch a way to understand the
remaining task regarding the origin of the small parameter
$\zeta_2-2/3$.

In Sect.~\ref{s:summary} we summarize past results that are necessary
for the present developments. In Sect.~\ref{s:ladder} we show how the
fusion rules can be used to determine the properties of the rungs in
the ladder diagrams appearing in the $n$th order correlation
functions. The first important new result is demonstrated in
Sect.~\ref{s:sanding} - the numerical coefficient contained in the
4-point rung is shown to be {\em small}, of the order of the anomaly
of $\zeta_2$. This result is crucial since it allows (to our knowledge
for the first time) to develop a perturbative calculation of the
anomalous parts of all the other exponents. The physical reason for
this result is that ( in Sect.~\ref{s:K41}) {\em we are developing the
theory around the K41 solution instead of the dissipative solution} as
was always attempted.  In Sect.~\ref{s:sanding} we pave the way for
the calculation of all the other exponents. In Sect.~\ref{s:surprise}
we calculate the scaling exponents by resumming the logarithmically
divergent ladder diagrams up to $O(\delta_2)$ (which is known in the
field-theoretic jargon as the ``1-loop order" in the renormalized
rungs). We find (admittedly to our surprise) that to this order the
scaling exponents are identical to K62. Like the latter they suffer
from the violation of the known requirement that $\zeta_n$ cannot
decrease with $n$ \cite{Fri}. In Sect.~\ref{s:2loop} we show that the
2-loop order cures the malaise of K62, and we present the result
(\ref{finalzn1}) for $\zeta_n$ that in our theoretical estimate is
valid for $n\le 12$. The exponent $\mu$ is also computed in this
Section. If one wanted results for $\zeta_n$ with higher values of $n$
one would need to go to 3-loop order (and see Sect.~\ref{s:discuss}
where the form of $\zeta_n$ to this order is presented), but the
present experimental situation does not warrant a theoretical
prediction of $\zeta_n$ for very high values of $n$.  In
Sect.~\ref{s:discuss} we summarize the paper, paying special attention
to the range of validity of the theory and to demonstrating that no
uncontrolled approximations were made.
\section{Summary of pertinent previous results}
\label{s:summary}                                                 
In this Section we present a brief summary of some past work which is
most pertinent. We refer to \cite{LP-1} as Paper I, to \cite{LP-2}
as Paper II and to \cite{96LP-3} as Paper III.
\subsection{The basic perturbation theory}
\label{ss:basic}
The starting point of the analysis are the Navier-Stokes equations for
the velocity field of an incompressible fluid with kinematic viscosity
$\nu$ which is forced by an external force ${\B. f}( {\B. r},t)$:
\def\OP{\raisebox{.2ex}{$\stackrel{\leftrightarrow}{\B. P}$}} 
\begin{equation} \left( { \partial \over \partial t} -
\nu\Delta^{2}\right) {\B. u} + \OP ( {\B. u} \cdot {\bbox{\nabla}})
{\B. u } = \OP {\B.  f}\,,  \label{b1}
\end{equation}
 where $\OP$ is the transverse projection operator $\OP \equiv -
\Delta^{-2} {\bbox{\nabla}}\times {\bbox{\nabla}}\times $. It is well
known, (and see for example
Paper I) that developing a perturbative approach
\cite{61Wyl,73MSR,LP-0} for the correlation functions and response
functions in terms of the Eulerian velocity ${\B. u}( {\B. r},t)$
results in a theory that is plagued with infra-red divergences. On the
other hand one can transform to new variables, and after the
transformation (which amounts to infinite partial resummations in the
perturbation theory) one finds a renormalized perturbation theory that
is finite, without any divergences in any order of the expansion
(cf.\cite{87BL} and Paper I) . One can achieve such a theory
using Lagrangian variables \cite{77Kra}; we find
it technically simpler to employ 
the Belinicher-L'vov transformation\cite{87BL},
 \begin{equation}
 {\B. v} [ {\B. r} _{0} | {\B. r},t] \equiv {\B. u}\big[ {\B. r}+
 {\bbox{\rho}}({\B. r}_{0},t),t)\big]\,,
 \label{b2}
 \end{equation}
 where  ${\bbox{\rho}}( {\B. r}_{0},t)$ is the Lagrangian trajectory of
 a fluid point which has started at point ${\B. r}
= {\B. r}_0$ at time $t=t_0$
 \begin{equation}
 {\bbox{\rho}}( {\B. r}_{0},t) =\int_0^t {\B. u}\big[ {\B. r}+
 {\bbox{\rho}}({\B. r}_{0},\tau),\tau\big] d\tau \ .
 \label{b3}
 \end{equation}
 The natural variables for a divergence free theory are the velocity
 {\it differences }
 \begin{equation}
 {\B. w}( {\B. r}_{0}| {\B. r},t)\equiv {\B. v}[ {\B. r}_{0}
 | {\B. r},t]- {\B. v}[ {\B. r}_{0} | {\B. r}_{0},t]\ .
 \label{b4}
 \end{equation}
 Since the averages of quantities that depend on one time only can be
computed at $t=0$, it follows that the average moments of these
BL-variables are the structure functions of the Eulerian field
$S_{n}(\B.R)$ defined by Eq.~(\ref{du1}).  It was shown \cite{87BL}
that these variables satisfy the Navier Stokes equations, and that one
can develop (cf. Paper I) a perturbation theory of the diagrammatic
type in which the natural quantities are the Green's function
$G_{\alpha\beta} ({\B. r}_{0} |{\B. r}, {\B. r}',t,t')$ and the
correlation function $F_{\alpha\beta} ({\B. r}_{0}
|{\B. r},{\B. r}',t,t')$:
 \begin{eqnarray}
 G _{\alpha\beta}( {\B. r}_{0}| {\B. r}, {\B. r}',t,t') &=&
 { \delta \langle w _{\alpha}( {\B. r}_{0} | {\B. r},t) \rangle
 \over  \delta f_{\beta}( {\B. r}',t')}{\Bigg |}_{f \to 0}\,,
 \label{b6} \\
 F_{\alpha\beta}( {\B. r}_{0}| {\B. r}, {\B. r}',t,t')
 &=&
 \langle w_{\alpha}( {\B. r}_{0}| {\B. r},t) w_{\beta}
 ( {\B. r}_{0} | {\B. r}',t')\rangle\ .
 \label{b7}
 \end{eqnarray}
Physically the Green's function is the mean response of the velocity
difference to the action of a vanishingly small forcing. In stationary
turbulence these quantities depend on $t'-t$ only, and we can denote
this time difference as $t$. The quantities satisfy the well known and
exact Dyson and Wyld coupled equations. The Dyson equation reads
 \begin{eqnarray}
 [{\partial \over  \partial  t}&-&\nu \Delta] G_{\alpha\beta}
 ( {\B. r}_{0}|{\B. r}, {\B. r}',t) = G^{0}_{\alpha\beta}
 ( {\B. r}_{0} | {\B. r}, {\B. r}',0^{+}) \delta (t)
 \nonumber \\
 &+&\int d {\B. r}_{2} G^{0}_{\alpha\delta}
 ({\B. r}_{0} | {\B. r}, {\B. r}_{2},0^{+}) \int d {\B. r}_{1}
 \int_ 0^t dt_{1}
  \label{b8}\\
 &\times&
 \Sigma_{\delta\gamma}( {\B. r}_{0}| {\B. r}_{2}, {\B. r}_{1} ,t_{1})
 G _{\gamma\beta}({\B. r}_{0}| {\B. r} _{1}, {\B. r}',t-t_{1})\,,
  \nonumber
 \end{eqnarray}
 where $ G^{0}_{\alpha\beta}( {\B. r}_{0} | {\B. r}, {\B. r}',0^{+})$ is
 the bare Green's function determined by Eq.~(3.20) in Paper I. The Wyld
equation has the form
 \begin{eqnarray}
 &&F_{\alpha\beta}( {\B. r}_{0} | {\B. r}, {\B. r}',t)
 = \int d {\B. r}_{1} d {\B. r}_{2}
 \int\limits_0^ \infty dt_{1} dt _{2} G_{\alpha\gamma}
 ( {\B. r}_{0} | {\B. r}, {\B. r}_1,t_{1})
 \nonumber\\
 &\times& \Big[ D_{\gamma\delta} ( {\B. r}_{1} \!-\! {\B.
 r}_{2},t\!-\!t_{1}+t_{2} ) + \Phi_{\gamma\delta}( {\B. r}_{0}|
 {\B. r}_{1},
 {\B. r}_{2}, t\!-\!t_{1}\!+\!t_{2}){\Big  ] }
 \nonumber\\
 &\times& G_{\delta\beta}( {\B. r}_{0} | {\B. r}',
 {\B. r} _{2} ,t_{2})\ .
  \label{b9}
 \end{eqnarray}
 In Eq. (\ref{b8}) the ``mass operator" $\Sigma$ is related to the
``eddy viscosity" whereas in Eq.~(\ref{b9}) the ``mass operator"
$\Phi$ is the renormalized ``nonlinear" noise which arises due to
turbulent excitations. Both these quantities are dependent on the
Green's function and the correlator, and thus the equations are
coupled.

  The main result of Paper I is a demonstration of the property of
  ``locality" in the Dyson and Wyld equations. This property means
  that given a value of $|{\B.  r}- {\B. r}_{0}|$ in Eq.~(\ref{b8}),
  the important contribution to the integral on the RHS comes from
  that region where $| {\B. r}_{1}- {\B.  r}_{0} |$ and $|
  {\B. r}_{2}- {\B. r}_{0}|$ are of the order of $| {\B.  r}- {\B. r}
  _{0}|$. In other words, all the integrals converge both in the upper
  and the lower limits. The same is true for the Wyld equation,
  meaning that in the limit of large $L$ and small $\eta$ these length
  scales disappear from the theory, and there is no natural cutoff in
  the integrals in the perturbative theory. In this case one cannot
  form a dimensionless parameter like $L/r$ or $r/\eta$ to carry
  dimensionless corrections to the K41 scaling exponents. For $ \eta
  \ll | {\B. r}- {\B.  r} _{0}| \ll L$ scale invariance prevails, and
  one finds precisely the K41 scaling exponents: 
\begin{eqnarray} G
  _{\alpha\beta} (\lambda {\B. r}_{0} | \lambda {\B. r},\lambda {\B.
  r}',\lambda ^{z} t) &=& \lambda ^{\beta_{2}} G _{\alpha\beta} (
  {\B. r} _{0}| {\B. r}, {\B. r}',t), \label{b10} \\
  F_{\alpha\beta}(\lambda {\B. r} _{0} | \lambda {\B. r},\lambda {\B.
  r}',\lambda { ^{z} }t) &=& \lambda ^{\zeta_{2}} F_{\alpha\beta}(
  {\B. r}_{0}| {\B. r}, {\B. r}',t)\ .  \nonumber \end{eqnarray} One
  can derive two scaling relations which hold order by order, {\em
  i.e.}  \begin{equation} 2z + \zeta_2=2\,,\quad z + 2\zeta_2=2 \ .
  \label{b11}
\end{equation}
The solution is $z = \zeta { _{2 } }= 2/3$. It was also shown that the
scaling exponent of the Green's function (\ref{b10}) is
$\beta { _{2} } = -3$. Extending such considerations to the higher order
structure functions leads to the order-by order K41 prediction that
$\zeta_n=n/3$.

Of course, the order by order result (\ref{b10}) which leads to
(\ref{b11}) is not necessarily the correct one.  If one could resum
all the diagrammatic expansion one could find nonperturbative answers
that may be different. The whole sum of diagrams may diverge when the
outer scale goes to infinity or the inner scale to zero, allowing a
renormalization scale to creep in even though the order-by-order
theory is convergent. The difficulty is that no one knows how to resum
the infinite expansion which exhibits no obvious small parameter.

In this paper we will propose a way out of this difficulty. The new
thinking is based on the fusion rules. Instead of considering fully
unfused correlation functions only, we will allow some coordinates to
be much closer together, say within a distance $\B.r$, whereas the rest
will be separated by a much larger distance, say of the order of $\B.R$
where $r\ll R$. We will show that we can form a dimensionless ratio
with $R/r$, and that such ratios {\em carry anomalous exponents that
are going to survive the process of fusion of coordinates in
correlation functions when we make structure functions}. We will thus
be able to recognize the anomalous exponents even though at first
sight there is no obvious renormalization scale.

To clarify how this mechanism works we need to remind ourselves of the
appearance of ladder diagrams in the theory of correlations
functions. Such diagrams appear in the most transparent way in
nonlinear Green's functions, and we review briefly our past results on
these objects.

 \subsection{ The Nonlinear Green's Functions}
\label{ss:NG}
The nonlinear Green's function $G_{2,2}( {\B. r}_{0} | x_{1} ,
x_{2},x_{3},x_{4})$ describes the response of a product of two
velocity fields taken at different space-time coordinates to the
action of two forces $\B.f$: 
\begin{eqnarray}
G_{2,2}^{\alpha\beta\gamma\delta}( {\B. r} { _{0} } | x { _{1} },x {
_{2} },x { _{3} },x { _{4} }) &=& \left\langle {\delta^2 w { _{\alpha}
}( {\B. r} { _{0} } | x { _{1} })w { _{\gamma} } ( {\B. r_0}|x_2)
\over \delta f { _{\beta} }( {\B. r} { _{0} } | x { _{3} })\delta f {
_{\delta} }( {\B. r} { _{0} } | x { _{4} }) }\right\rangle\,,
\nonumber\\ &&\label{c1} 
\end{eqnarray} 
where for brevity we use the notation $x_{j} \equiv \{ {\B. r} { _{j}
},t { _{j} }\} $. Similarly, one defines the nonlinear Green's
functions $\B.G_{p,p}$ as the response of the product of $p$ velocity
differences between $p$ distinct points and $\B.r_0$ to the action of
$p$ forces in different points.

In a Gaussian theory (which ours is not) $\B.G_{2,2}$ would be the
products of the linear Green's functions like $G { ^{\alpha\beta} }(
{\B. r} { _{0} } | x { _{1} },x { _{3} })G { ^{\gamma\delta} }(
{\B. r} { _{0} } | x { _{2} },x { _{4} })$. In a non-Gaussian theory
it is natural to assume that this quantity is a homogeneous function
of its arguments when they are in the scaling regime.  This means that
\begin{eqnarray} &&G_{2,2 }^{\alpha\beta\gamma\delta}( {\B. r} { _{0}
} | {\lambda}{\B. r} { _{1} },{\lambda}{ ^{z} }t { _{1}
},{\lambda}{\B.  r} { _{2} },{\lambda}{ ^{z} }t { _{2}
},{\lambda}{\B. r} { _{3} },{\lambda}{ ^{z} }t { _{3} },
{\lambda}{\B. r} { _{4} },{\lambda}{ ^{z} }t { _{4} }) \nonumber \\
&=& {\lambda}^{\beta_{4} }G_{2}^{\alpha\beta\gamma\delta} ( {\B. r} {
_{0} } | x { _{1} },x { _{2} },x { _{3} },x { _{4} }) \label{c2}
\end{eqnarray}

From the Gaussian decomposition of this quantity we would guess that
$\b { _{4} } = 2\b { _{2} } = -6$.  The proof of locality in Paper I
means that there is no perturbative mechanism to change this scaling
index. On the other hand, this quantity, which is a function of four
space-time coordinates $x { _{i} }$ has scaling properties that are
not exhausted by the overall scaling exponent $\b _4$. We have
shown in Paper II that when we consider its dependence on {\it ratios} of
space-time coordinates in their asymptotic regimes w e pick up a set of
anomalous scaling exponents.  The main result of Paper II was that in
the regime $r_1 \sim r _2 \ll r { _{3 } } \sim r { _{4 } }$ the
diagrammatic expansion of this object produces logarithms like ln$(r {
_{3} }/r { _{1} })$ to some power. It was explained that the sum of
such logarithmically large contributions is given by $(r {_{3} }/r {
_{1} }) { ^{\Delta} }$ with some anomalous exponent $\Delta$. To make
the appearance of anomalous exponents evident we review the simplest
object that resums to logarithms, {\em i.e.}  the series of ``ladder
diagrams".
\begin{figure}
\epsfxsize=8.6truecm
\epsfbox{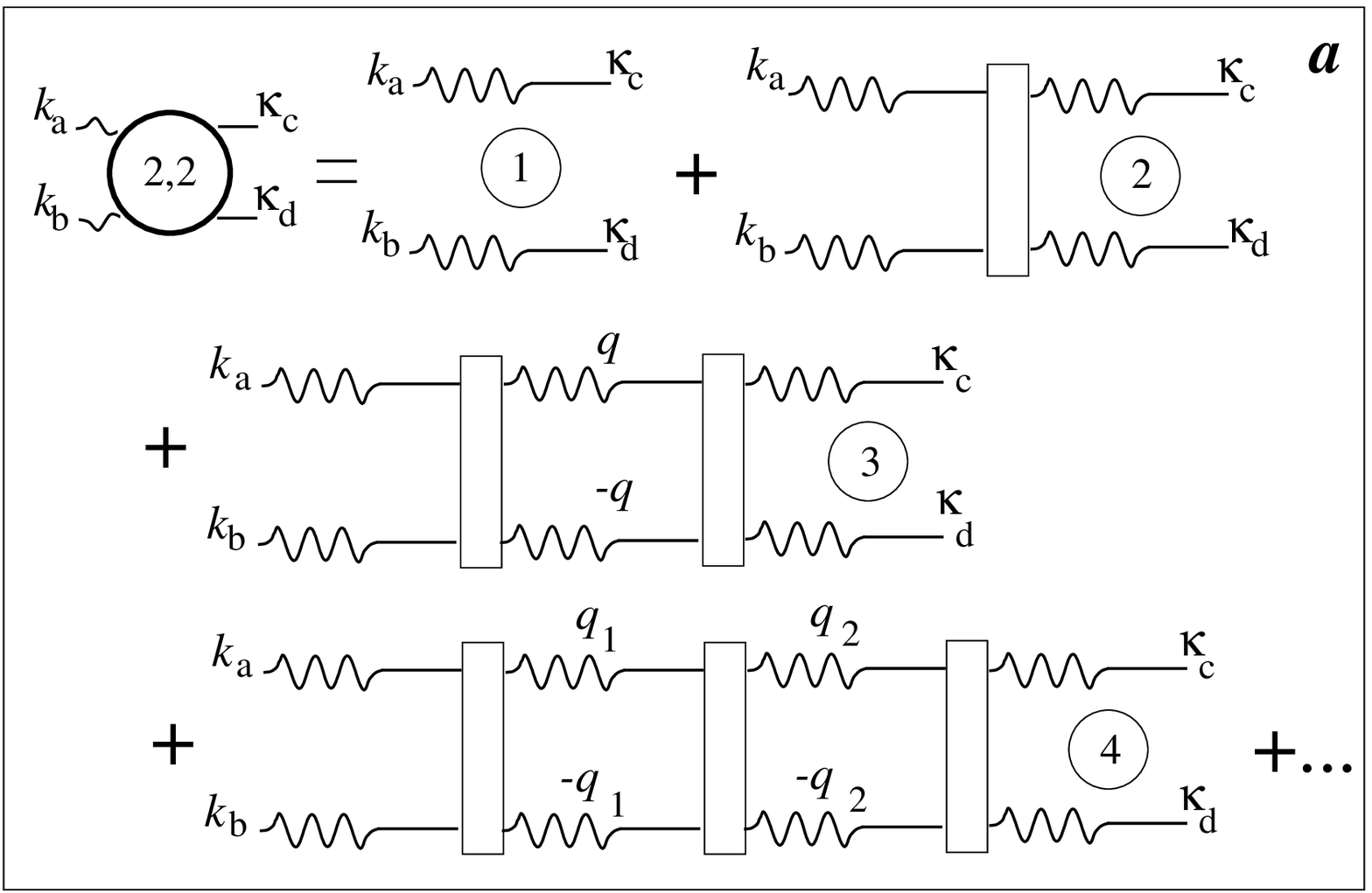}
\epsfxsize=8.6truecm
\epsfbox{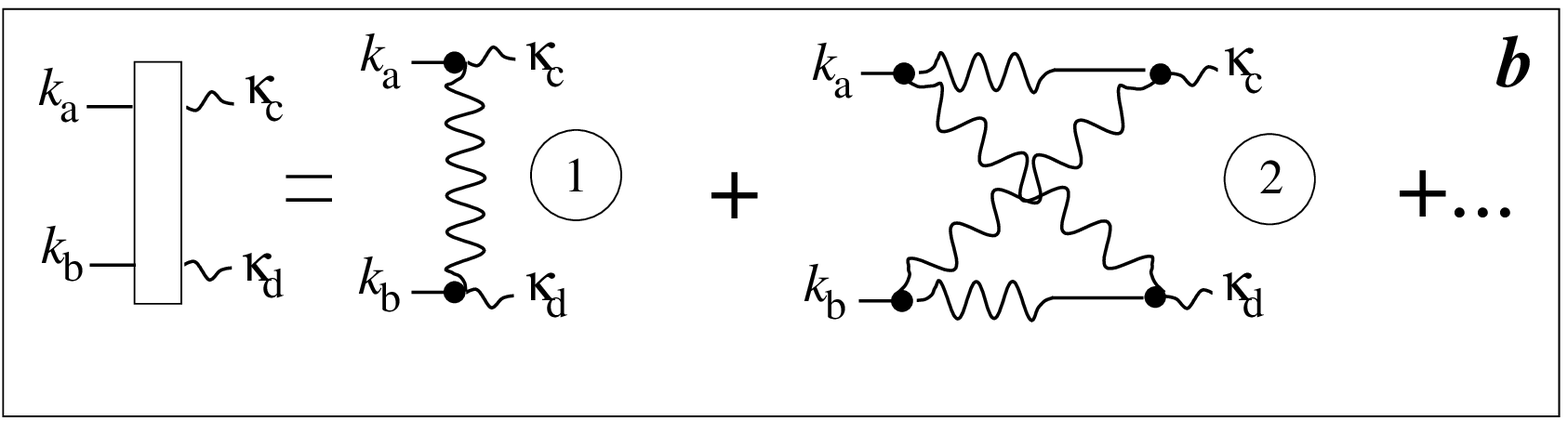}
\narrowtext
\vskip 0.2cm
\caption{Diagrams for  ${\bf G}_{2,2}
({\bf k}_a,\omega_a,{\bf k}_b,\omega_b,
{\bf \kappa}_c,\omega_c,{\bf \kappa}_d,\omega_d)$ (Panel (a)) and of the
rung ${\bf R}({\bf  k}_a, {\bf k}_b,{\bf \kappa}_c,{\bf \kappa}_d)$
(Panel (b)). Diagram (1) in Panel (a) is the Gaussian contribution made of
a product of two linear Green's functions. Diagram (2) is the skeleton
contribution, and diagrams (3) and (4) are the 1-loop and 2-loop
contributions respectively. In Panel (b) we show the beginning of the
infinite series expansion for the rung, with diagram (1) being the
bare rung.}
\label{fig-1}
\end{figure}
\begin{figure}
\epsfxsize=8.6truecm
\epsfbox{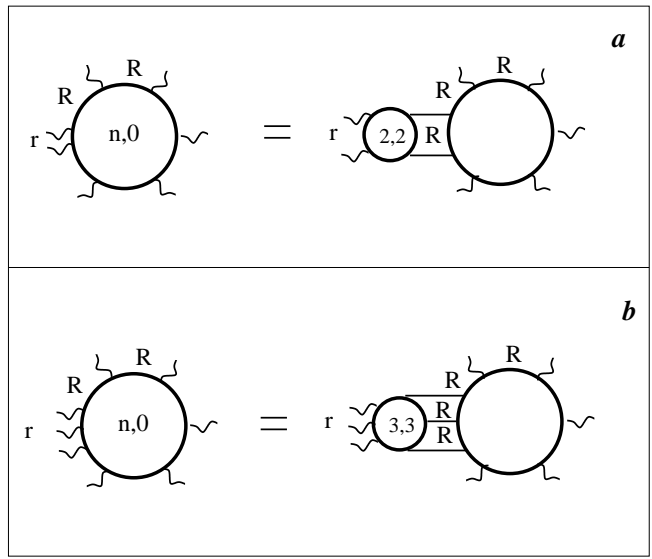}
\vskip 0.2cm 
\caption{A diagrammatic representation of the fusion process. A typical
$n$ the order correlation function is represented by $n$ wavy lines
decorating a circle. Panel (a): The fusion of 2 coordinates to within
a distance $r$ which is much smaller that the typical separations $R$
between the other coordinates. Panel (b): The fusion of three
coordinates to within a distance $r\ll R $ from each other. The
availability of all the possible diagrams afforded by the topological
rules allows us to expose a contribution that is precisely $G_{p,p}$
with $p=2$ and 3 respectively.}
\label{fig-2}
\end{figure}
The diagrammatic representation of the nonlinear Green's function
(\ref{c1}) is shown in Fig.~\ref{fig-1}a, where the notation of the
diagrams is explained shortly in the figure legend and at length in
Appendix~\ref{ap-A}.  The 4-point ``rungs" appearing in the ladders were
represented in Paper II as an infinite series expansion whose
beginning is shown in Fig.~\ref{fig-1}b. The rest of the expansion
involves exceedingly complicated diagrams that we failed to resum
analytically. Nevertheless, just from general properties one could
show that the ladder with $n$ rungs contains a contribution of order
$\big[ \Delta \ln ( r { _{3} }/r {_{1}}) \big]^{n}/n!$. The summation
of all these contributions gives a term proportional to $(r _{3} /r {
_{1} })^{\Delta}$, and this is the observation that we want to build
on in this paper.
\section{Ladder diagrams in $n$'th order correlation functions}
\label{s:ladder}
In this Section we demonstrate how the anomalous exponents of higher
order correlation functions can be related to resummed ladder
diagrams. The idea is to consider a typical $n$th order correlation
function and to almost fuse $p$ coordinates, $p<n$, chosen from
the available $n$
coordinates.  The point to observe is that the diagrammatic theory
allows us to write, upon observation, all the topologically possible
diagrams appearing in the expansion of a given object. Thus for
example consider Fig.~\ref{fig-2} where we represent a general $n$th
order correlation function, in which two coordinates are a distance
$r$ from each other, and all the rest are a distance $R$ from them and
from each other. While coalescing the two coordinates we pull out a
contribution that reads $G_{2,2}$ connected to the rest of the diagram
with $n-2$ coordinates. We know that this contribution exists since it
is allowed topologically. But this fragment is represented in its turn
by the sum of the ladder diagrams that we present in Fig.~\ref{fig-1}.
We thus state that whenever we are about to fuse two coordinates in
any $n$th order correlation function we can expose a series of
diagrams that are the same as those that appear in the expansion of
the nonlinear Green's function $G_{2,2}( {\B. r}_{0} | x_{1} ,
x_{2},x_{3},x_{4})$, together with the logarithmic divergences that
are associated with them. In Appendix~\ref{ap-A} we explain that in
doing so we really take into account all the necessary contributions,
leaving nothing uncontrolled.

Now we employ the fusion rules. These tell us that {\em as a function
of the two fusing coordinates} the $n$th order correlation function is a
homogeneous function whose homogeneity exponent is
$\zeta_2$. Accordingly, if we succeed to resum the ladder diagrams in
the limit $r\ll R$ we should find an anomalous part of $\zeta_2$ from
the dimensionless power $\left(R/r\right)^\Delta$.

Similarly, we can almost fuse 3, 4 or $p$ coordinates, and accordingly
pull out of the diagram for the $n$th order correlation functions more
complex ladders with 3, 4 or $p$ struts. For example in
Fig.~\ref{fig-2}b we show how the fusion of 3 coordinates singles
out a fragment that is $G_{3,3}$ whose ladder diagrams have three
struts cf. Fig.~\ref{fig-3}. The fusion rules guarantee that the $n$th
order correlation is a homogeneous function of the $p$ fusing
coordinates with $\zeta_p$ being the homogeneity exponent.  We will
show that these more complex ladders also resum into power laws in
$R/r$, being responsible for the anomalous parts of $\zeta_p$.

At this point all this is a bit formal, since we do not have an
explicit form of the rungs in the ladder diagrams, and we can compute
nothing without this knowledge. In the next two Sections we will
address this issue and demonstrate that a judicious use of the fusion
rules dictates enough knowledge of the rungs to take us through
a useful calculation.

\section{Building the theory on the background of K41}
\label{s:K41}
In this Section we reorganize the theory such that Kolmogorov's 41
theory serves as its ``free" limit. In other words, we aim at
achieving a theory in which resummations of divergent contributions
would directly give the {\em anomalous} parts of the scaling
exponents; the K41 parts should be obvious order-by-order. This is
done in two steps, that are correspondingly presented in
Subsects.~\ref{ss:resum} and \ref{ss:renorm}.

\subsection{Resummation into K41 propagators}
\label{ss:resum}
It was explained in Subsect.~\ref{ss:basic} that our theory is
developed in the BL-representation, to eliminate spurious IR
divergences that stem from the sweeping interactions. The main result
of Paper I was that after line-resummation each diagram in the
BL-diagrammatic expansion of the propagators (Green's function and
double correlation function) converged in the infrared and the
ultraviolet regimes. Accordingly, K41 scaling is a solution of the order-by-order
theory. Nevertheless, the propagators in the BL-representation lose
translational invariance, and are therefore not diagonal in Fourier
space. For the purpose of actual calculations it is extremely
advantageous to rearrange the theory such that the BL-propagators
become again diagonal in Fourier space.

The actual resummation that is necessary is presented in
Appendix~\ref{K41prop}.  It results in a diagrammatic theory that is
topologically exactly the same as the standard Wyld diagrammatic
expansion before line resummation. There are two differences as
explained in the Appendix~\ref{K41prop}. For the purposes of our
considerations below the main issue is the simple form of the
the propagators that
appear as lines in the diagrams: they exhibit K41 scaling exponents:
\begin{eqnarray} \label{GK41}
G_{\alpha\beta}(\B.k,\omega)
&=&P_{\alpha\beta}(\B.k) g(k,\omega), \ 
g(k,\omega)=\frac{1}{\omega+i\gamma(k)}\ ,\\
F_{\alpha\beta}(\B.k,\omega)&=&P_{\alpha\beta}(\B.k) f(k,\omega),\ 
f(k,\omega)=\frac{\phi(k)}{\omega^2+\gamma^2(k)} . \label{FK41}
\end{eqnarray}
In these formulae the scaling exponents are carried by
\begin{equation}
\gamma(k)=c_\gamma \bar\epsilon^{1/3} k^{2/3}\ ,
\qquad \phi(k)=c_\phi\bar\epsilon k^{-3} \
,\label{gamK41}\\
\end{equation}
where $c_\gamma$ and $c_\phi$ are dimensionless constants.
\subsection{Renormalization to K41 4-point rung}
\label{ss:renorm}
In this Subsection we determine the form of the 4-point rungs of the ladder
diagrams in two steps. These two steps are based on the following observation: 
the diagrammatic
expansion of the rung includes many diagrams, some of which contain in them
additional subsets of ladder diagrams. In the first step we will consider the rungs 
as if all the diagrams appearing in their infinite
series were resummed, {\em except} for their own internal subsets of ladder
diagrams. In the second step we will consider also the ladder diagrams
appearing in the series for the rung. We aim to a situation in
which all the ladders that appear in the theory, like in Fig.~\ref{fig-1}, contain already
renormalized rungs. However, instead of evaluating the rungs from
actual resummations we are going to determine their form using the
fusion rules. Thus in the first step we find the form of
the rung that results, upon fusion, in K41 scaling exponents.  In the
second step we recognize that the rungs themselves have ladders,
leading to an anomalous correction in the scaling properties of the
rungs themselves. This being accomplished, we will have our final form
of the rung. Then we turn to the ladder diagrams appearing in the fused
correlation functions, using the rung as a basic building block of the theory. 
All anomalies of
all the measurable statistical objects will result from resummations of the remaining
ladder diagrams.

Consider Fig.~\ref{fig-1}a, in which the rung appears as an object. It
is given in terms of an infinite series of diagrams in
Fig.~\ref{fig-1}b. It is in fact a 4-point vertex depending on four
$\B.k$-vectors and four frequencies. As a first step we consider the
value of the rung when all the frequencies are zero, denoting it in
this limit as $\B.R(\B.k_a,\B.k_b,\B.\kappa_c,\B.\kappa_d)$. At a
later point we will explain that this is sufficient for our
purposes. The bare value of this object can be read directly from
diagram (1) in Fig.~\ref{fig-1}b, with two bare BL-vertices
$\B.\Gamma$ and one double correlation function. The answer is
\begin{eqnarray}
&&R^{\alpha\beta\gamma\delta}_0(\B.k_a,\B.k_b,\B.
\kappa_c,\B.\kappa_d)\nonumber\\
&&=\d_0\bar\epsilon^{1/3}
\frac{\Gamma^{\alpha\gamma\sigma}(\B.k_a,\B.\kappa_c,\B.k_e)
\Gamma^{\beta\delta\sigma}(\B.k_b,\B.\kappa_d,-\B.k_e)}{k_e^{13/3}} \ ,
\label{rung}
\end{eqnarray}
where $\B.k_e\equiv \B.k_a-\B.\kappa_c=\B.\kappa_d-\B.k_b$, and
$\d_0$ is a dimensionless constant.

We demonstrate now that if we use this bare form of the rung the
fusion rules would predict {\em dissipative} exponents, $\zeta_n=n$.
We first demonstrate this in the context of $\zeta_2$. Consider a
general $n$ order correlation function as in Fig.~\ref{fig-2}a, and
fuse two coordinates, pulling out the fragment of $\B.G_{2,2}$ as
shown on the RHS of Fig.~\ref{fig-2}a.  We will now compute the
scaling exponent by finding the $r$ dependence of this fragment when
the two coordinates approach each other to a {\em small} distance
$r\ll R$ where $R$ is the typical distance between all the other
coordinates. To find the $r$ dependence we must integrate according to
the explanation in Appendix~\ref{ap-self}, and to this aim we
introduce the object $T_2(r,\B.\kappa)$:
\begin{eqnarray}
&&T_2(r,\B.\kappa) = \int\frac{d\B.k_a}{(2\pi)^3}
4\sin^2(\frac{1}{2}\B.k_a\cdot\B.r)\nonumber\\
&&\times\int \frac{d\omega_a}{2\pi} G_{2,2}
(\B.k_a,\omega_a,-\B.k_a,-\omega_a,\B.\kappa,0,-\B.\kappa,0) \ .
\label{T2def}
\end{eqnarray}
We are interested in the $r$ dependence of this object in the limit
$r\kappa\ll 1$. To calculate $T_2(r,\B.\kappa)$ in this limit we
return to Fig.~\ref{fig-1}a. Obviously the Gaussian contribution
diagram (1) is irrelevant in this limit. The skeleton diagram (2)
contributes the following integral:
\begin{eqnarray}
&&T^{\rm s}_2(r,\B.\kappa) \approx \int
\frac{d\omega_a}{2\pi}\int\frac{d\B.k_a}{(2\pi)^3}
4\sin^2(\frac{1}{2}\B.k_a\cdot\B.r)\nonumber\\
&&\times g (k_a,\omega_a)g(-k_a,-\omega_a)
R_0({\B.k_a,-\B.k_a,\B.\kappa,-\B.\kappa})\ ,
\label{T2s}
\end{eqnarray}
where the tensor indices of the rung were contracted for the
longitudinal contribution.  The superscript ``s" is used here and
below to denote skeleton contributions. We note that in the limit
$k\gg\kappa$ the BL vertices are proportional to the smallest
wavevector $\kappa$. Thus the rung is proportional to
$\kappa^2/k_a^{13/3}$. Integrating over the frequencies of the two
Green's functions $g(k_a,\omega_a)$ in this rung [cf.
Eqs.~(\ref{GK41}, \ref{gamK41})] results in the evaluation
$1/[\gamma(k_a) k_a^{13/3}]\propto k_a^{-5}$. Thus the $r$ dependence
of $T_2^{\rm s}$ is given by
\begin{eqnarray}
&&T^{\rm s}_2(r,\B.\kappa) \propto \int\frac{d\B.k_a}{(2\pi)^3}
4\sin^2(\frac{1}{2}\B.k_a\cdot\B.r)\frac{1}{k_a^5} \ .
\label{T2ss}
\end{eqnarray}
Up to logarithmic corrections this integral is proportional to $r^2$
which is the dissipative solution. Similarly, if we use the bare rung
in the diagram in Fig.~\ref{fig-1}b to determine $\zeta_3$ we will
find $\zeta_3=3$. In general we will find $\zeta_n=n$ instead of the
K41 value of $n/3$. Now one could think that the correct values of the
inertial range exponents may be obtained from resumming all the ladder
diagrams with the bare rungs. This was the point of view proposed in
Paper I.  In such a case the sought after correction to the
scaling exponents is of the order of unity, and it is unclear how to
develop a controlled resummation.  In this paper we point out a new
way, based on the existence of a renormalized rung which gives, upon
fusion, K41 exponents {\em before} ladder resummations.  The
characteristics of the renormalized rung in such a scheme are dictated
by the fusion rules.

Next we want to determine the {\em renormalized} form of the rung. To
this aim we repeat the exercise of integrating over the two Green's
functions and the rung, with the vertices determined as before in the
limit $k_a\approx k_b\gg \kappa_c\approx \kappa_d$. But now we leave the
exponent of $k_a$ in the asymptotic evaluation of the rung free, and
{\em demand} that the result of the integration will be $k_a^{-5/3}$.
We find that this requires that $R_0\propto k_a^{-3}$.

We are now in a position to propose a renormalized form of the rung
which in a proper calculation could be obtained by summing up all the
non-ladder diagrams that contribute to this rung. This conforms with
our basic hypothesis that all non-ladder diagrams contribute to K41,
whereas the ladders are responsible for the anomalous scaling. Since
K41 does not allow $L$ renormalization we propose the form
\begin{eqnarray}
&&R^{\alpha\beta\gamma\delta}(\B.k_a,\B.k_b,\B.\kappa_c,
\B.\kappa_d)\nonumber\\
&&=\d\bar\epsilon^{1/3} \frac{\Gamma^{\alpha\gamma\sigma}
(\B.k_a,\B.\kappa_c,\B.k_e)
\Gamma^{\beta\delta\sigma}(\B.k_b,\B.\kappa_d,-\B.k_e)}
{k_e^{3} \kappa_c^{2/3}\kappa_d^{2/3}}
\ . \label{renorung}
\end{eqnarray}
This form gives the K41 overall scaling exponent (which is the same as
in the bare rung $\B.R_0$, and in addition agrees with the fusion
rules for the second order correlation function with a 
K41 scaling exponent. In addition it is
symmetric, as it should be, for exchanging the indices $a$ and $b$
together with $c$ and $d$. Note that $\delta$ is now a renormalized
unknown dimensionless parameter which will be determined later.

To proceed, we note that our actual calculation (see below) depends
really only on the asymptotic properties of the rung, which are
rigidly determined by the fusion rules. We can thus attempt to
simplify the form of the rung as much as possible, preserving the
asymptotic and parity properties unchanged. In particular we note that
the BL vertices $\B.\Gamma(\B.k_a,\B.\kappa_c,\B.k_e)$ have
complicated structure which makes calculations involving them rather
difficult. Therefore we propose to use instead Eulerian vertices
$\B.V(\B.k_a,\B.\kappa_c,\B.k_e)$ corrected by a
factor $-2(\B.k_b\cdot\B.\kappa_c)/[k_a^2+k_b^2+\kappa_c^2]$. The
correction is aimed at reproducing the asymptotic behavior of the BL
vertex $\Gamma^{\alpha\gamma\sigma}(\B.k_a,\B.\kappa_c,\B.k_e)\sim
{\rm min}
\{k_a,k_b,\kappa_c\}$.
Thus instead of (\ref{renorung}) one has
\begin{eqnarray}
&&R^{\alpha\beta\gamma\delta}(\B.k_a,\B.k_b,\B.\kappa_c,\B.\kappa_d)
={-4\, \d\,
\bar\epsilon^{1/3}(\B.\kappa_c\cdot\B.k_e)(\B.\kappa_d\cdot\B.k_ e)
\over [k_a^2+\kappa_c^2+k_e^2][k_b^2+\kappa_d^2+k_e^2]}\nonumber\\
&\times& \frac{ V^{\alpha\gamma\sigma}(\B.k_a,\B.\kappa_c,\B.k_e)
V^{\beta\delta\sigma}(\B.k_b,\B.\kappa_d,-\B.k_e)}{k_e^{3} (\kappa_c
\kappa_d)^{2/3}}
\ . \label{rung2}
\end{eqnarray}
As a further simplification of the actual calculations we will use a
1-dimensional reduction of the problem (preserving the asymptotic
scaling properties and parity) in which instead of 3-dimensional
integrations $\int d^3k/(2\pi)^3$ we will use 1-dimensional one
$\int_{-\infty}^\infty dk/2\pi$.  Then we can disregard the vector
indices and replace $V^{\alpha\gamma\sigma}(\B.k_a,\B.\kappa_c,
\B.k_e)\rightarrow k_a$, $(\B.k\cdot\B.k')\rightarrow kk'$ (keeping
the signs) and $k_e^3$ in the denominator by $|k_e|$ (because we
replaced 3-d by 1-d integration). The 1-dimensional version of the
rung (\ref{rung2}) turns into
\begin{equation}
R(k_a,k_b,\kappa_c,\kappa_d) ={-4 \,\d\, 
\bar\epsilon^{1/3}k_ak_b \kappa_c
\kappa_d|k_e|\over
[k_a^2+\kappa_c^2+k_e^2][k_b^2+\kappa_d^2+k_e^2]|
\kappa_c \kappa_d|^{2/3}}
\ . \label{rung3}
\end{equation}

Note that here $k_a,k_b,\kappa_c,\kappa_d$ are in the interval $\pm
\infty$ and that they carry signs in order to preserve the parity of
the rungs.  $k_a,k_b$ are incoming wave vectors and
$\kappa_c,\kappa_d$ are outgoing, and they conserve momentum,
\begin{equation}
k_a+k_b=\kappa_c+\kappa_d \ . \label{conserve}
\end{equation}
Substituting into (\ref{rung3}) $k_e=k_a-\kappa_c=k_b-\kappa_d$ one
gets finally:
\BEA.rung4
&&R(k_a,k_b,\kappa_c,\kappa_d) \br
&=&-
\d\,\bar\e ^{1/3} {k_ak_b \kappa_c \kappa_d |k_a-\kappa_c|\over
 [k_a^2-k_a \kappa_c+\kappa_c^2][k_b^2-k_b \kappa_d+
\kappa_d^2]|\kappa_c \kappa_d|^{2/3}} \ .
\EEA
In particular,
\begin{eqnarray}
&& R(k,-k,\kappa,-\kappa)= {- \d\,\bar\e ^{1/3} k^2 \kappa^{2/3}
|k-\kappa|\over
 [k^2-k \kappa+\kappa^2]^2}, \label{rung5}\\
&&R(k,-k,\kappa,\kappa')=
{\d\,\bar\e ^{1/3}{\rm sign}(\kappa\,\kappa')
 \, |\kappa\,\kappa'|^{1/3} \over
|k|  }\,\nonumber\\ &&\quad\quad \quad \mbox{for}
\ \kappa,\kappa'\ll k\ .
\label{rung6}
\end{eqnarray}

To check that we get the right K41 scaling exponents with the new
renormalized rung we need to recalculate the 1-dimensional version of
Eq.~(\ref{T2s}) with (\ref{rung4}) for the rung:
\begin{eqnarray}
T_2^{\rm s}(r,\kappa)&=&\int_{-\infty}^\infty\frac{dk}
{2\pi}4\sin^2\Big(\frac{kr}{2}\Big)
R(k,-k,\kappa,-\kappa)\nonumber\\
&\times&\int_{-\infty}^\infty\frac{d\omega}
{2\pi}g(k,\omega)g(-k,-\omega) \
. \label{T2s1d}
\end{eqnarray}
Using Eq.~(\ref{GK41}) the frequency integral yields
$-1/2\gamma(k)$. Thus
\begin{equation}
T_2^{\rm s}(r,\kappa)=\frac{\delta\kappa^{2/3}}{c_\gamma \pi}
\int_{-\infty}^\infty dk \frac{|k|^{4/3}
|k-\kappa|\sin^2\Big(\frac{kr}{2}\Big)}
{[k^2-k\kappa +\kappa^2]^2} \ .
\end{equation}
In the limit $\kappa\to 0$ the integral simplifies to
\begin{equation}
T_2^{\rm s}(r,\kappa)=2\tilde \delta\kappa^{2/3}
\int_0^\infty \frac{dk}{k^{5/3}} \sin^2
\Big(\frac{kr}{2}\Big) \ , \label{T2use}
\end{equation}
where
\begin{equation}
\tilde\delta\equiv \frac{\delta}{\pi c_\gamma}
\ . \label{tildeldef}
\end{equation}
This integral is elementary, reading
\begin{equation}
T_2^{\rm s}(r,\kappa)=-\frac{1}{2}\tilde\delta (\kappa r)^{2/3}
\Gamma\Big(-\frac{2}{3}\Big)\approx 2 \tilde
\delta (\kappa r)^{2/3} \ ,
\end{equation}
where $\Gamma(x)$ is the gamma-function.

The point to notice is that in the asymptotic limit $\kappa\to 0$ the
only properties of the rung that guaranteed the appearance of the
scaling exponent 2/3 are the asymptotic properties that we preserved
in the series of simplifications leading to (\ref{rung4}). In general,
we will show below that the series of simplifications of the model
form of the rung are of absolutely no import also for the calculation
of the {\em anomalous} scaling exponents up to 1-loop order. We will
show below that we get {\em precisely} the same exponents in this
order with any arbitrary analytic form of the rung, with tensor
indices or without, in 3-d form or 1-d form or whatever, as long as
the asymptotics are preserved, as they are. In 2-loop order this is no
longer true. The actual numbers obtained in the 2-loop order are model
dependent.  We will show however that the sensitivity of the predicted
exponents $\zeta_n$ to the model for the rung is small as long as
$n<8$.  We need the 2-loop order mainly to make sure that it corrects
for some unacceptable properties of the 1-loop results for higher
order correlation functions.
\begin{figure}
\epsfxsize=8.1 truecm
\epsfbox{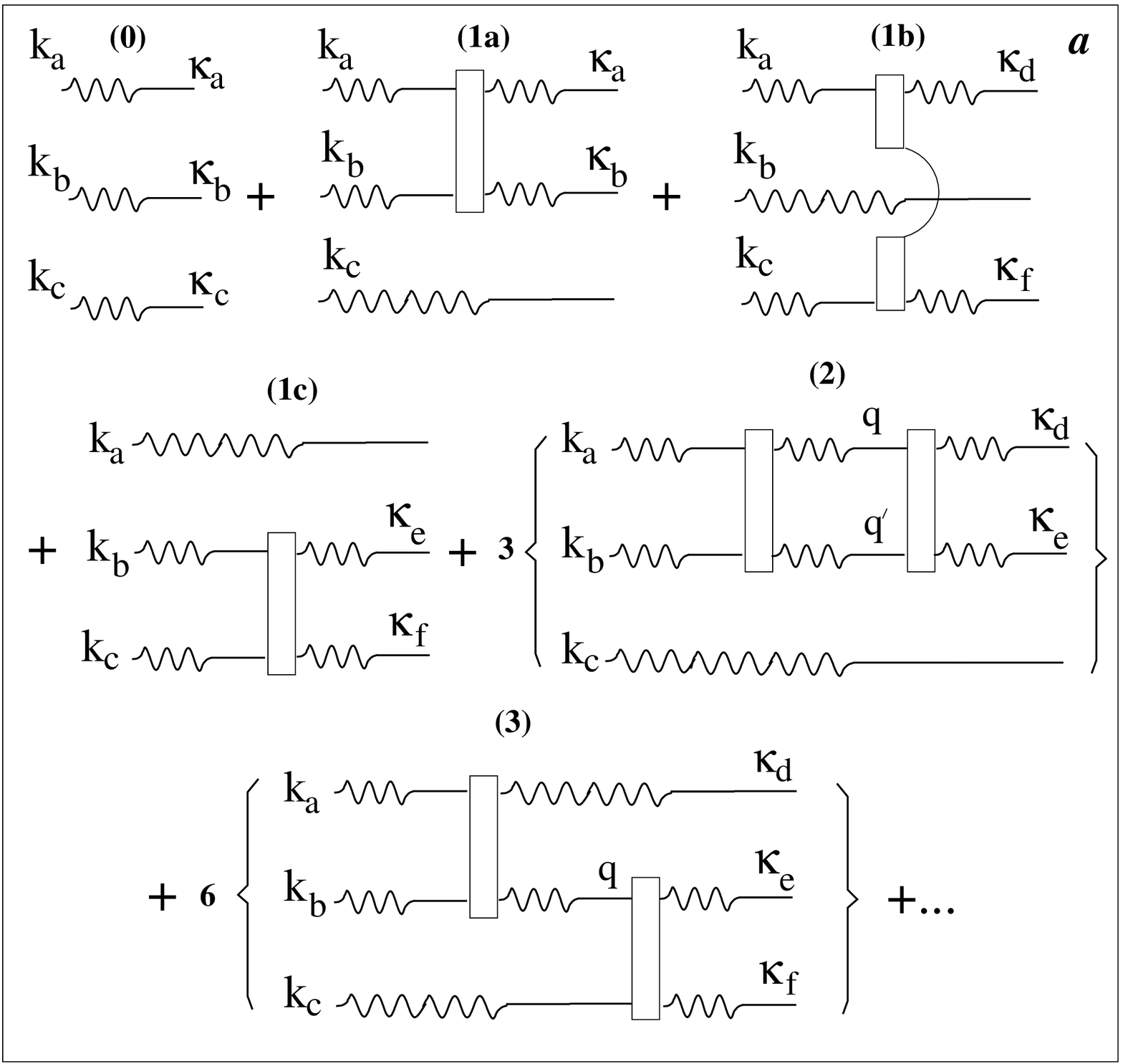}
\epsfxsize=8.1 truecm
\epsfbox{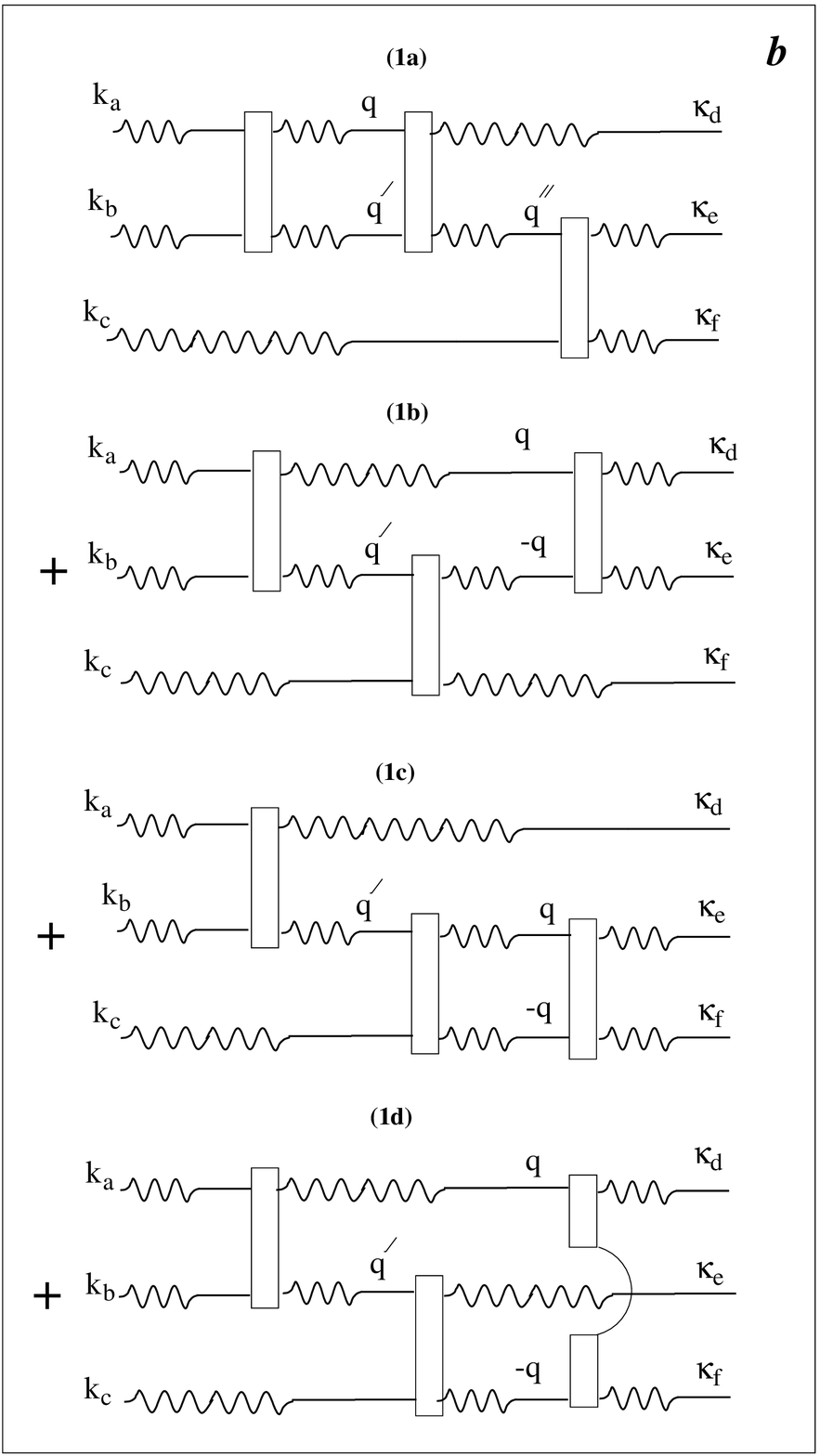}
\caption{Diagrammatic expansion of ${\bf G}_{3,3}$  (panel a):
Contributions with no rungs, (diagram 0), one rung and two rungs.
(panel b): Contributions with three rungs.}
\label{fig-3}
\end{figure}
Before we proceed we need to check the self consistency of our
approach.  We need to make sure that all higher order non-linear
Green's function $\B.G_{p,p}$ (the response of $p$ velocities to $p$
forcing) yield, upon fusion, the correct K41 exponent for $p$th order
correlation functions $\zeta^{^{\rm K41}}_p=p/3$. For this aim we have
to consider the so called {\em skeleton} diagrams which are the lowest
order connected diagrams without loops.  For $\B.G_{2,2}$ this is
diagram (2) in Fig.~\ref{fig-1}a.  for $\B.G_{3,3}$ the skeleton
contribution are shown as diagrams (3) on Fig.~\ref{fig-3}a.  We must
make sure that the skeleton diagrams, upon fusion, yield K41 scaling
$\zeta^{^{\rm K41}}_p=p/3$ for the appropriate $p^{\rm th}$-order
correlation functions, since our grand hypothesis is that the
anomalous scaling comes only from ladder resummations.

This test of self consistency is presented in Appendix~\ref{ap-self}.
The important conclusion of this Appendix is that the skeleton
 diagrams for $G_{p,p}(\{k_j,\kappa_j\})$ (with asymptotics of the
 rung defined by the two-point fusion rules with $\z^{^{\rm
 K41}}_2=2/3$) automatically reproduces the K41 scaling
 exponent $\Z.p   = p/3  $ when $p$ points are fused.

\section{Sanding the floor in the one-loop order}
\label{s:sanding}
In this Section we demonstrate the most important new property of the
resummed theory, {\em i.e.} that the rungs in the ladder diagrams
appear with a small parameter. This will allow us to develop a {\em
controlled} ladder resummation, something that to our knowledge has
never been available before. In fact, we will lay out in this Section
all that is needed to calculate the scaling exponents in the 1-loop
order.  In Subsect.~\ref{ss:small} we demonstrate that the prefactor
$\delta$ of the rung is the order of $\delta_2$ which is the anomalous
part of $\zeta_2$ and thus small. In Subsect.~\ref{ss:anomal} we
consider the anomalous exponent of the rung itself, denoted as
$\delta_a$, and stemming from ladder resummations within the rung
infinite series representation. In Subsect.~\ref{ss:contr} we
reconsider the contribution of the skeleton diagrams to the scaling
exponents upon fusion, taking into account the anomaly of the rung.
In Subsect.~\ref{ss:corrs} we throw in the inputs: the fact that
$\zeta_3=1$ and the experimental value of $\zeta_2$. The result is
Eq.~(\ref{allsame}) which states that in the 1-loop order all the
unknown parameters are numerically identical. From this point the
calculation of all the other scaling exponents in the 1-loop order is
straightforward.

\subsection{The 4-point rung is small!}
\label{ss:small}
Here we show that the coefficient $\delta$ in front of the
renormalized 4-point rung (\ref{renorung}) is of the order of the
correction to K41 of the scaling exponent $\z_2$:
\BE.delta
\d_2=\z_2-\z^{^{\rm   K41}}_2\approx 0.03\ .
\EE
To this aim consider the 1-dimensional version of the quantity
$T_2(r,\B.\kappa)$ of
(\ref{T2def}):
\begin{eqnarray}
&&T_2(r,\kappa) = \int\limits_{-\infty}^\infty\frac{dk_a}{2\pi}
4\sin^2(\frac{1}{2}k_a r)\nonumber\\
&&\times \int\limits_{-\infty}^\infty \frac{d\omega_a}{2\pi}G_{2,2}
(k_a,\omega_a,-k_a,-\omega_a,\kappa,0,-\kappa,0) \ .
\label{T2def1d}
\end{eqnarray}
We will examine the ratio of the contributions of the 1-loop diagram
(3), denoted below as $T_2^{(1)}(r,\kappa)$, to the contribution of
the skeleton diagram Eq.~(\ref{T2s1d}) [diagram (2) in
Fig.~\ref{fig-1}a]. After performing the frequency integrals the
1-dimensional form of $T_2^{(1)}(r,\kappa)$ [cf. diagram (3) in
Fig.~\ref{fig-1}a]   is:
\begin{eqnarray}
&&T^{(1)}_2(r,\kappa) = 
\int\limits_{-\infty}^\infty\frac{dk_a}{\pi\gamma(k_a)}
\sin^2(\frac{1}{2}k_a r)\nonumber\\
&&\times \int\limits_{-\infty}^\infty
\frac{dq}{2\pi\gamma(k_a)}R(k_a,-k_a,q,-q)
R(q,-q,\kappa,-\kappa) \ .
\label{T211d}
\end{eqnarray}
In the asymptotic limit defined by $\kappa r\to 0$ the main
contribution to the integral comes from the 2 symmetric regions of the
q-integration in which $|\kappa|\ll |q| \ll |k_a|$. In these regions
we use the form (\ref{rung6}).  We calculate
\begin{equation}
T^{(1)}_2(r,\kappa) \approx 2\tilde\delta^2 \kappa^{2/3}
\int\limits_0^\infty\frac{dk_a}{k_a^{5/3}}\sin^2(\frac{1}{2}k_ar)
\int\limits_{\kappa}^{k_a} \frac{dq}{q} \ .
\label{T21final}
\end{equation}
As expected the loop integral over $q$ produces a logarithmic
contribution.  At this point we use the asymptotical identity
\begin{equation}
\lim_{\kappa r \to 0}\int_0^\infty d(kr)f(kr)
 \ln\Big(\frac{k}{\kappa}\Big)
=\ln\Big(\frac{1}{r\kappa}\Big)\int_0^\infty d(kr)f(kr) \ ,
 \label{identity}
\end{equation}
which produces, upon comparison with Eq.~(\ref{T2use}) the final result
\begin{equation}
T_2^{(1)} (r,\kappa)=   \tilde \d\, \ln \Big[\frac{1}{\kappa r}\Big]  \,
 T_2^{(\rm s)} (r,\kappa) \ . \label{2nd-1}
\end{equation}
The factor $\tilde \delta$ that was introduced in
Eq.~(\ref{tildeldef}) reappears here in front of the logarithm as the
effective parameter of expansion.

Analogously one computes the leading contribution of the two-loop
diagram (4) in Fig.~\ref{fig-1}a. This is done explicitly in
Subsect. \ref{ss:2l2} and Appendix~\ref{2lint}:
\begin{equation}
T_2^{(2)} (r,\kappa)= \frac{1}{2}
\Big[\tilde \d \ln \Big(\frac{1}{r\kappa}\Big) \Big]^2   T_2^{(\rm s)}
(r,\kappa) \ , \label{2nd-2}
\EE
and, in general the leading contribution of the $n$-loop diagram:
\BE.2nd-n
T_2^{(n)} (r,\kappa)= \frac{1}{n!}
\Big[\tilde \d \ln \Big(\frac{1}{r\kappa}\Big) \Big]^n   T_2^{(\rm
s)}(r,\kappa)
\ .
\EE
The sum of all these contributions is as follows:
\BEA.2nd-sum
T_2(r,\kappa)&=&T_2^{(\rm s)} (r,\kappa) +\sum_{m=1}^\infty T_2^{(m)}
(r,\kappa) \br
&=&T_2^{(\rm s)} (r,\kappa) \sum_{m=0}^\infty  \frac{1}{m!}
\Big[\tilde \d \ln \Big(\frac{1}{r\kappa}\Big) \Big]^m =
 \frac{ T_2^{(\rm s)} (r,\kappa)}{(r\kappa)^{ \tilde\d }}   \ .
\EEA
We see that, as usual, resummation of the leading contributions from
the logarithmic ladder diagrams results in the power function with the
exponent $\tilde \d$ which is the prefactor of the logarithm in the
one-loop diagram, see Eq.~(\ref{2nd-1}). Because the expected
correction $\d_2$ to the K41 exponent $\Z.2 $ is small ( $\d_2\approx
0.03$ ) we conclude that the prefactor of the rung $\tilde \d$ is
small as well! This allows us to begin with the one-loop approximation
in computing the higher order scaling exponents $\z_p$ with $p>2$.

\subsection{Anomalous correction of the rung asymptotics}
\label{ss:anomal}
So far we have disregarded the explicit appearance of ladder diagrams
in the infinite series that defines the rung itself. As pointed in
Paper II the same kind of ladder resummation that is responsible for
the anomaly of the exponents of the nonlinear Green's functions will
also contribute an anomalous part to the scaling properties of the
rung.  Nevertheless the outer and inner scale do not appear in
the rung either, and therefore the anomaly is explicit only in the asymptotic
regime where we have a ratio of large and small scales. In this
Subsection we flush out this anomaly.

Instead of (\ref{rung6}) we expect
\BEA.rung7
R_{\rm a} (k,-k,\kappa,\kappa')=
{\d\,\bar\e ^{1/3}{\rm sign}(\kappa\,\kappa') \,
|\kappa\,\kappa'|^{1/3+\d_{\rm  a}    }
\over
|k|^{1+2\d_{\rm  a}}  } \\ \nn
\qquad \qquad  \qquad \qquad\qquad \qquad  \qquad \qquad
\mbox{for} \ k\gg \kappa,\kappa'\,,
\EEA
with some anomalous exponent $ \d_{\rm a} $ which is expected (and
later demonstrated) to be
of the order of $\tilde\d$ as it stems from the same origin.
This correction to the asymptotics may be achieved, for
example by the following model form of the rung (\ref{rung4}):
\BE.rung8
R_{\rm a} (k_ a,k_ b,\kappa_c,\kappa_d )
=R(k_a,k_ b,\kappa_c\kappa_d )\Bigg(
\frac{| \kappa_c\kappa_d| }{|k_a-\kappa_c|^2}
\Bigg)^{\d_{\rm a}}.
\EE
As before, we will argue that the exact analytic form of the rung is
not important for our calculations, and only the asymptotic scaling
form is essential. This statement will be shown to be exact in the
1-loop order. We thus need at this point only to preserve the
essential properties, {\em i.e.} that the outer scale $L$ cannot
appear due to locality, that the rung has to be symmetric with respect
$a,b\to c,d$, etc.

\subsection{Contributions of the skeleton diagrams with
 the anomalous 4-point rung}
\label{ss:contr}
In this Subsection we reconsider the skeleton diagrams appearing in
the nonlinear Green's functions $G_{p,p}$ but taking into account the
anomaly of the rung. In other words we are going to compute the
scaling exponents accounting only for the ladder resummation {\em
inside} the rung, but not the ladder resummation with the anomalous
renormalized rungs. This final step will be done in
Sects.~\ref{s:surprise} and \ref{s:2loop}.

We are interested in the scaling exponents of structure functions in
$r$ representation, and these
are obtained from the correlation functions in $k$ representation as detailed in
Appendix~\ref{ap-self}. Upon fusion we obtain automatically contributions
behaving like nonliner Green's functions. Accordingly the objects of
interest in the analysis below are the nonlinear Green's functions in which
the $k$ dependence of the fusing coordinates is transformed to $r$
representation. The outgoing wavevectors $\kappa$ are left as are, and the
outgoing frequencis can be put to zero with impunity. This
results in objects defined in a mixed $r,\kappa$ representation, which
we denote as $ T_p(r,\{\kappa_j'\})$:
\begin{eqnarray}\nn
 T_p(r,\{\kappa_j'\})&=&\Int \prod_{i=1}^p\frac{d\, \o_i d\,k_i}{(2\pi)^2}
\d(\o_1+\dots \o_p)\d(k_1+\dots k_p) \\
&\times& f_p(r,\{k_j \})G_{p,p}(\{k_j,\omega_j,\kappa_j,0\})\ .
 \label{Tpdef}
\EEA
Here $f_p(r,\{k_j \})$ are one-dimensional versions of the functions
$f_p(\B.r,\{\B.k_j \})$ defined by Eq.~(\ref{str}). The set
$\{\kappa_j\}$ denotes all the outgoing wavevectors.

We consider the skeleton contributions to the nonlinear Green's
function $G_{p,p}$, denoted by $G^{\rm s}_{p,p}$. Similarly to the
definition (\ref{T2s1d}) we introduce $T_p^{\rm s}(r,\{\kappa_j\})$ as
\begin{eqnarray}\nn
 T^{\rm s}_p(r,\{k_j'\})&=&\Int \prod_{i=1}^p\frac{d\, \o_i d\,k_i}{(2\pi)^2}
\d(\o_1+\dots \o_p)\d(k_1+\dots k_p) \\
&\times& f_p(r,\{k_j \})G^{\rm s}_{p,p}(\{k_j,\omega_j,\kappa_j,0\})\ .
\label{Tpsdef}
\EEA
Repeating the calculation of Appendix~\ref{ap-self} in the asymptotic
regime $\kappa_j\,r\ll 1$ but with the redefined rung (\ref{rung7})
one gets
\BE.Tpr
T_p(r,\{\kappa_j\})=C_p (\bar\e r )^{p/3}\,r^{p\,  \d_{\rm a}}
\prod_{j=1}^p |\kappa_j|^{1/3+\d_{\rm a}} \ .
\EE
Here $C_p$ are dimensionless constants that absorb all the numerical
factors. In fact, the result (\ref{Tpr}) could be guessed directly by
recognizing that every rung which is connected with the ``outgoing''
Green's functions $G(\kappa_j,0)$ contributes to $T_p(r,\{\kappa_j\})$
a factor $|\k_j|^{1/3+\d_{\rm a}}$. All together they give
$\prod_{j=1}^p |\kappa_j|^{1/3+\d_{\rm a}}$.  Convergence of the
integrals over $\kappa_j$ and $\omega_j$ implies that neither inner
nor outer scales may appear, and therefore dimensional consideration
require a factor $r^{p(\case{1}{3}+\d_{\rm a})}$.

\subsection{The 2nd and 3rd order correlation functions: relations between
$\d_{\rm a}$, $\d_2$ and $\tilde\d$ } 
\label{ss:corrs}
Consider first the scaling exponent $\z_2$. In Subsect.~\ref{ss:small}
we showed that the resummation of the ladder diagrams leads to an
anomalous correction to the exponent $\Z.2 = \case{2}{3}$ which is
$-\tilde \d$ (cf.  Eq.~(\ref{2nd-sum}). But according to
Eq.~(\ref{Tpr}) the ladder in the skeleton contribution brings in an
additional correction $2\,
\d_{\rm a}$.  Altogether we have in the 1-loop order
\begin{equation}
\z_2=\case{2}{3}-\tilde \d+2 \, \d_{\rm a} \ . \label{z21loop}
\end{equation}
Therefore the exponent
$\d_2$ defined by (\ref{delta}) may be expressed as follows:
\BE.rel1
\d_2=2\, \d_{\rm a}- \tilde \d \ .
\EE

Another relation between the exponents will follow from the analysis
of the fusion of three points. To 1-loop order the nonlinear Green's
function $G_{3,3}$ has the skeleton contribution diagram 3 in
Fig.~\ref{fig-3}a, and the 1-loop diagrams in Fig.~\ref{fig-3}b. The
skeleton contribution can be read directly from Eq.~(\ref{Tpr}):
\BE.Tpr3
T^{\rm s}_3(r,\{\kappa_j\})=C_3 \bar\e r \,r^{3\,  \d_{\rm a}}
 |\kappa_1\kappa_2\kappa_3|^{1/3+\d_{\rm a}} \ .
\EE
To discuss the other contributions we refer to Fig.~\ref{fig-3} in
which all the diagram of $G_{3,3}$ with zero, one and two rungs are
represented. We have one diagram with no rung, three with one, nine
with 2. The multiplicity of 3 in the diagrams of type (2) represent
the three possible connections of two struts by two rungs. The
multiplicity of 6 in the diagrams of type (3) represent the different
pair-permutations of three struts. In general there are $3^n$ diagrams
with $n$ rungs, out of which three will have one disconnected
strut. Diagrams with disconnected struts will not contribute in the
asymptotic regime that interests us here. Thus out of the diagrams in
Fig.~\ref{fig-3}a only the skeleton diagram (3) remains in the
asymptotic regime.

In general, with $n$ rungs we have $3^n-3$ fully linked diagrams. This
number is $6 [(3^{n-1}-1)/2]$, and the number $[(3^{n-1}-1)/2]$ counts
the topologically distinct fully linked diagrams with $n$ rungs. Thus
for example we represent in Fig.~\ref{fig-3}b the four topologically
distinct contributions with three rungs.  These are all the 1-loop
ladder diagrams contributing to the 3'rd order correlation
function. We show now that of these four terms diagram (1a) does not
contribute a logarithmic divergence, whereas the other three
contribute the same logarithmic term. In fact, this is the beginning
of a systematic rule: the only diagrams that contribute logarithmic
terms in the 1-loop order are those in which the last rung appears to
the right of the skeleton diagrams. Similar rules will be established
below for higher loop contributions.

Consider then the 1-loop diagram (1a) in Fig.~\ref{fig-3}b in which
this rule is not obeyed. We focus on the loop made by the two rungs
and the Green's functions $q$ and $q'$, considering the asymptotic
regime $k_a,k_b\gg q \gg
\kappa_a, \kappa_b$. This is the only regime in which a 
logarithmic divergence is possible. In this regime $q'\approx
q''\approx k_a+k_b$. Thus the rung $R_a(k_a,k_b,q,q')$ contributes to
the loop $q^{1/3+\delta_a}$. The rung $R_{q,q',\kappa_a,q''}$ and the
Green's function $G(q')$ do not contribute any $q$ dependence to the
integrand. The $\omega$ integration over the product of the Green's
functions $G(q)$ and $G(q')$ gives approximately $1/\gamma(q')$ and
again contributes no $q$ dependence. Finally we have the evaluation
\begin{equation}
T_3^{(1a)}\propto \int_\kappa^{k_a}dq q^{1/3+\delta_a} \ .
\end{equation}
Clearly, this diagram does not exhibit a logarithmic divergence and as
such it does not contribute to the renormalization of the scaling
exponent.

The other three diagrams in Fig.~\ref{fig-3}b (namely 1b,1c and 1d)
are different, they all
have a logarithmic divergence. The reason for the difference is that
in these three diagrams there are four Green's functions, instead of
five in diagram (1a), which carry large wavevectors. This is the same
situation as in the skeleton diagram (3) in Fig.~\ref{fig-1}a. In the
loop we have now two Green's functions, instead of one in diagram
(1a), that carry small wavevectors $q$. This difference leads to a
different $q$ dependence in the loop, and to a logarithmic
divergence. We demonstrate this explicitly in the next paragraph, but
we already draw the conclusion which is general: 1-loop ladder
diagrams with logarithmic divergences are those in which the
additional rung (compared to skeleton diagram) has been positioned to
the {\em right} of the skeleton structure.

Explicitly, consider diagram (1b) in Fig.~\ref{fig-3}b. The rung
$R_a(k_a,k_b,q,q')$ contributes $q^{1/3+\delta_a}$ as before. But now
also the rung $R_a(q',k_c,-q,\kappa_c)$ contributes the same
$q$-dependence. On the other hand the rung
$R_a(q,-q,\kappa_a,\kappa_b)$ contributes $|q|^{-1-2\delta_a}$. The
$\omega$ integration with the product of the two Green's functions
$G(q,\omega)G(-q,-\omega)$ is the same as (\ref{T2s1d}) leading to
$-1/2\gamma(q)$. In total we have a logarithmic integral.

The diagram (1d) is very similar to (1b); it has the same rung
structure at the left, and the rightmost rung is
$R_a(q,-q,\kappa_a,\kappa_c)$.  This makes no difference to the $q$
dependence and thus to the logarithmic divergence or to the factor in
front of the logarithm.  Diagram (1c) is slightly different, having
the third rung on the same ladder as the second rung. Nevertheless the
rung $R_a(q',k_c,q,-q))$ contributes exactly the same $q$ dependence
as the {\em two } rungs in diagrams (1b) or (1d). Thus it yields at
the end the same factor with the same logarithm. Finally, comparing 1c
to diagram 3 in Fig.~\ref{fig-1}a we see that the loop structures are
identical in both, and thus if diagram 3 had a prefactor
$\tilde\delta$, we can immediately conclude that the {\em three}
diagrams (1b, 1c and 1d) will result in a total prefactor of $3\tilde
\delta$:
\begin{equation}
T_3^{(1)}(r,\{\kappa_j\})=  3\tilde \d\, \ln
 \Big[\frac{1}{r\,\kappa}\Big]  \,
T_3^{(s)}(r,\{\kappa_j\}) \ , \label{T31final}
\end{equation}
where $\kappa\equiv [\kappa_1\kappa_2\kappa_3]^{1/3}$. The leading
contribution from the higher loop diagrams can be seen to contribute
higher order terms in the series of a power law, similarly to the
mechanism displayed in Eqs.~(\ref{2nd-1}--\ref{2nd-sum}):
\begin{equation}
T_3(r,\{\kappa_j\})=   \frac{T_3^{(s)}(r,\{\kappa_j\})}
{[r\kappa]^{3\tilde
\d}} \,.
\end{equation}
Substituting (\ref{Tpr3}) we find finally
\begin{equation}
T_3(r,\{\kappa_j\})=   C_3\bar \epsilon
 [r\kappa]^{1+3\delta_a-3\tilde \delta}
 \,.
\end{equation}
Accordingly to 1-loop order we write
\begin{equation}
\zeta_3=1+3\delta_a-3\tilde \delta \ .
\end{equation}
At this point we use the exact, nonperturbative result that $\zeta_3=1$
to find the relationship between $\delta_a$ and $\tilde\delta$:~~$
\delta_a=\tilde\delta $.
Together with (\ref{rel1}) we get the important conclusion that all
our $\delta$'s are the same:
\begin{equation}
\delta_2=\delta_a=\tilde\delta=\zeta_2-\Z.2  \approx 0.03 \ .
 \label{allsame}
\end{equation}
We should stress that this important result is obtained using only
the asymptotic scaling properties of the rung. Changing the explicit
form of the rung without ruining the asymptotics will affect only
the subleading terms in the analysis. The leading logarithmic terms
are insensitive to the details of the analytic form of the rung.
\begin{figure}
\epsfxsize=8.1truecm
\epsfbox{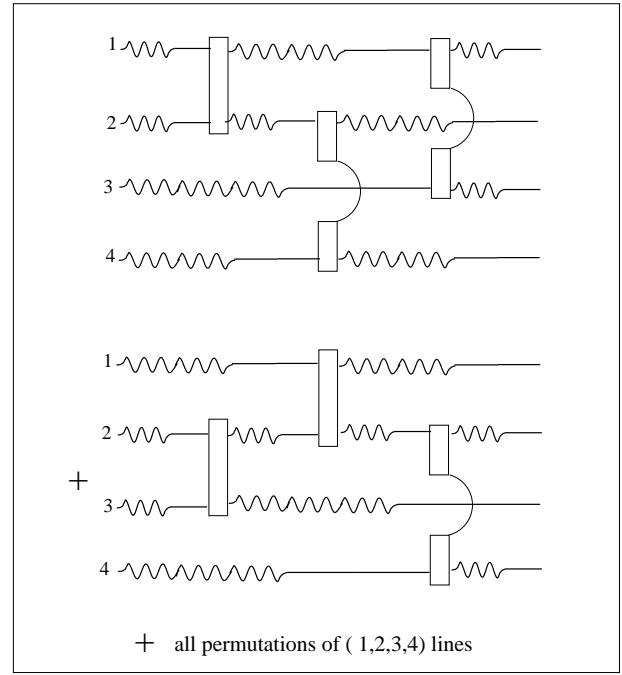}
\vskip 0.3 cm
\caption{The skeleton contributions in the diagrammatic
 expansion of ${\bf G}_{4,4}$.}
\label{fig-4}
\end{figure}
\section{anomalous scaling exponents in the 1-loop approximation:
 Surprise,   surprise}
\label{s:surprise}
We are poised to compute now the anomalous corrections to all the
scaling exponents of the $p$-order correlation functions in the 1-loop
approximation. Start with the 4th order nonlinear Green's function,
and consider the skeleton diagrams in Fig.~\ref{fig-4}. In the one
loop order, to obtain a logarithmic divergence in the asymptotic
regime we must add the additional rung {\em on the right} of the
skeleton structure.  The combinatorics are elementary: Each skeleton
diagram can host a new rung on the right in six different ways. Once a
rung has been put in place the leading (logarithmic) contribution to
the loop integral is the same as the loop integrals considered in the
last Section. It gives the same logarithm {\em with the same
prefactor}. The only difference is in the combinatorics. We can thus
write by inspection
\begin{equation}
T_4^{(1)}(r,\{\kappa_j\})=  6\tilde \d\, \ln
\Big[\frac{1}{r\,\kappa}\Big]  \,
T_4^{(s)}(r,\{\kappa_j\}) \ ,
\end{equation}
where $\kappa$ is the geometric mean of all the $\kappa_j$.
Resumming the leading contributions of the  higher order loop diagrams
results in the power law
\begin{equation}
T_4(r,\{\kappa_j\})=   \frac{T_4^{(s)}(r,\{\kappa_j\})}{[r\kappa]^{6\tilde
\d}} \,.
\end{equation}
Using Eq.~(\ref{Tpr}) we reach the final result
\begin{equation}
\zeta_4 = 4/3+4\delta_a-6\tilde\delta = 4/3-2\delta_2 \ , \quad
\mbox{1-loop order.}
\end{equation}

The analysis of the 1-loop order contribution to the anomalous
exponents of the $p$-order correlation functions is as
straightforward. There are $p(p-1)/2$ possibilities to append an
additional rung to the right of the skeleton structure of the $p$-order
nonlinear Green's function. All these diagrams contribute identical
leading order logarithmic terms, with the same prefactor, summing up
to an anomalous correction to the scaling exponent of the skeleton
diagrams which is $\tilde \delta p(p-1)/2$. According to
Eq.~(\ref{Tpr}) the scaling exponent of the skeleton contribution 
itself is corrected with
respect to K41 by $p\delta_a$.  Thus altogether
\begin{equation}
\zeta_p=\frac{p}{3}+p\delta_a-\tilde \delta \frac{p(p-1)}{2}  \ .
\end{equation}
Using Eq.~(\ref{allsame})
\begin{equation}
\label{K62again}
\zeta_p=\frac{p}{3}-\delta_2 \frac{p(p-3)}{2}  \ , 
\quad \mbox{1-loop order}.
\end{equation}
We note that this formula, which is valid in our case to 1-loop order
only, is identical in prediction to Kolmogorov's log-normal
phenomenological model (known as K62). This is interesting, as it
stems from the nontrivial topology of the ladder diagrams, in which
only the most leading were considered. The present authors find the
connection between lognormality and ladder diagrams unexpected.

Nevertheless we should recognize that in the present approach this
result has a limited region of validity. The analysis of the 2-loop
order which is provided below will show that Eq.~(\ref{K62}) is only valid
when $p\delta_2 \ll 1$. The 2-loop order will contribute {\em positive}
terms of the order of $\delta_2^2 p^2(p-3)$, {\em reducing} the negative
tendency of the correction to K41. Accordingly the present theory will
not suffer from the well know deficiencies of the K62 log-normal model 
which for us is only a first order result.
\section{Anomalous Scaling Exponents in the 2-loop approximation: K62
is cured}
\label{s:2loop}
In this Section we calculate the 2-loop contributions to the scaling
exponents $\zeta_p$. Even though these contributions are very small
when $p\delta_2$ is small (for, say, $p<6$), they become important for
larger values of $p$ where K62 begins to turn down the $n$ dependence
of $\zeta_n$. In addition this
calculation allows to present clear ranges of validity for the 1-loop 
and 2-loop calculations. 
\subsection{2-loop contributions to $\zeta_2$}
\label{ss:2l2}
We consider the 2-loop diagram (4) in Fig.~\ref{fig-1}a. Substituting
it instead of $G_{2,2}$ in Eq.~(\ref{T2def1d}) we obtain the quantity
$T_2^{(2)}(r,\kappa)$.  We want to compute the correction that this
diagram gives to the skeleton diagram (2), and to this aim we divide
it by $T_2^{\rm s}(r,\kappa)$. In the asymptotic regime $\kappa r\ll 1$ the
loop integrals over $q_1$ and $q_2$ contribute mostly in the range
$k\gg q_1,q_2\gg \kappa$. In this regime the integrals over
$k_a,\omega_a$ cancel from the ratio of
$T_2^{(2)}(r,\kappa)/T_2^{\rm s}(r,\kappa)$.  In addition the Green's
functions $G(\kappa_c)$ and $G(\kappa_d)$ cancel.  Thus this ratio can
be read from the ratio of the corresponding diagrams for $G_{2,2}$, 
or taking the diagram (4) and
amputating the incoming and outgoing legs. We still need to divide by
the rung in diagram (2) where $k_a$ is replaced by $1/r$:
\begin{eqnarray}
&&\frac{T_2^{(2)}(r,\kappa)}{T_2^{\rm s}(r,\kappa)}
=\frac{1}{R(1/r,-1/r,\kappa,-\kappa)}
\int\limits_{-\infty}^{\infty} \frac{dq_1 dq_2 J(q_1)\,J(q_2)}
{(2\pi) q_1 q_2 }
 \nonumber\\
&&\times R(r^{-1}\! ,\!-r^{-1}\!,q_1,\!-q_1)
R(q_1,\!-q_1,q_2,\!-q_2)R(q_2,\!-q_2,
\kappa,\! -\kappa),
\nonumber \\ &&J(q)=\int_{-\infty}^\infty
\frac{d\omega}{2\pi} G(q,\omega)G(-q,-\omega)=
-\frac{1}{2\gamma(q)} \ . \label{Kkp}
\end{eqnarray}
In Appendix~\ref{2lint} we analyze this integral in the asymptotic
limit $\kappa r\ll 1$ with the final result
\begin{equation}
T_2^{(2)}(r,\kappa)=\tilde
\delta^2\left[\frac{1}{2}\ln^2{\Big(\frac{1}
{\kappa r}\Big)}+b_1\ln{\Big(\frac{1}{\kappa
r}\Big)}\right] T_2^{\rm s}(r,\kappa)
\ ,
\label{Kkp1}
\end{equation}
where $b_1$ is a dimensionless constant
\begin{equation}
b_1\approx -0.434 \ . \label{cb1}
\end{equation}
In Eq.~(\ref{Kkp1}) the $\ln^2$ term accounts for the exponentiation
of the 1-loop contribution, whereas the $\ln$ term provides the 2-loop
correction to the scaling exponent $\zeta_2$. Instead of
Eq.~(\ref{rel1}) we now read
\begin{equation}
\delta_2=2\delta_a-\tilde \delta-b_1 \tilde \delta^2 \ .
\label{z22loop}
\end{equation}
A second relation between these exponents will be derived in the next
subsection.
\subsection{2-loop contributions to $\zeta_3$}
\label{ss:2l3}
The calculation to $O(\delta_2^2)$ of the contributions to $\zeta_3$
and of higher order $\zeta_p$ 
due to ladder resummations introduces for the first time 6-point
irreducible interactions amplitudes.  These appear in the ladder diagrams as
rungs with six legs, arising from diagrams that due to their
topology cannot be resummed into reducible contributions consisting 
of two 4-point rungs and one Green's function. The
6-point rung is discussed in Appendices~\ref{ap-A} and \ref{6rung}. In particular
in Appendix \ref{6rung} we explain why the {\em functional dependence} of 
$\zeta_p$ on $p$ can be understood completely on the basis of the analysis
of ladders with 4-point rungs. This stems from the fact that the reducible
and irreducible contributions to the 6-point
rung are of the same order, and their combinatorical factors are identical. 

There are many possible two loop diagrams involving 4-point rungs that
appear in the expansion
of $G_{3,3}$. However, we are only interested in those contributing a
logarithmic divergence in the asymptotic regime. As before, to get the
relevant diagrams we need to append the last rung {\em to the right}
of the existing 1-loop structure. Thus, we begin with the 3
logarithmic diagrams in Fig.~\ref{fig-3}b ({\em i.e.}  1b, 1c and 1d)
and consider all the diagrams that are obtained by adding an additional rung
on the right which connects two struts. The nine resulting diagrams
are shown in Appendix~\ref{ap-nine}. These diagrams are subdivided
into two groups: three diagrams in which the last rung connects the
same struts as the previous rung, and six diagrams in which the last
rung connects different struts. In Appendix~\ref{ap-nine} we explain
that {\em the first group of diagrams gives exactly the same
asymptotic integral as the 2-loop contribution to $\zeta_2$}. This
statement should be reiterated because of its importance to the
structure of the theory: the integrals are different, but once the
limits $k\gg q_1,q_2\gg \kappa$ are taken, the resulting integrals
coalesce with those computed in the previous Subsection. Thus the
contribution to $\zeta_3$ from these three diagrams will be $3b_1
\tilde \delta^2$ (which is 3 times larger than the corresponding
contribution to $\zeta_2$).

The six diagrams of the second group look topologically different, but
again in the asymptotic regime coalesce into an identical integral
Eq.~(\ref{canonical}) with $\tilde\Psi\to \tilde \Psi_2$, where
\begin{eqnarray}\label{tpsi2}
&&\tilde \Psi_2(q_1,q_2)\\ \nn
&=& { q_1 |q_1|^{1/3}|q_1+q_2|^{4/3}{\rm sign} (q_2)
\over (|q_1|^{2/3}+ |q_2|^{2/3}+|q_1+q_2|^{2/3})
(q_1^2+q_2q_1+q_2^2)} \ .
\end{eqnarray}
Following the procedure outlined in Appendix~\ref{extract} we find the
coefficients of expansion
\begin{equation}
a_2=1 \ , \quad b_2\approx -0.55 \ . \label{cb2}
\end{equation}
Finally we get the 2-loop form of $\zeta_3$:
\begin{equation}
\zeta_3=1+3\delta_a-3\tilde \delta
-\tilde\delta^2(3b_1+6b_2) \ . \label{2loopz3}
\end{equation}
Demanding again $\zeta_3=1$ we find from
Eqs.~(\ref{z22loop}, \ref{2loopz3}):
\begin{eqnarray}
\tilde\delta &=&\delta_2-(b_1+4b_2)\delta_2^2\
+O(\delta_2^3), \label{imp1}\\
\delta_a&=&\tilde \delta [1+(b_1+2b_2)\tilde \delta]
 \ . \label{imp2}
\end{eqnarray}
These results are used in the next Subsection to calculate $\zeta_n$
for $n\ge 3$ to 2-loop order.
\subsection{2-loop contributions to $\zeta_p$, $p\ge 4$}
\label{ss:2l4}
The calculation of the contribution of 4-point rungs to $\zeta_p$ 
for higher values of $p$ does
not necessitate the evaluation of new integrals. In Appendix~\ref{ap-nine}
we explain that all the 2-loop integrals appearing in the ladders of
$G_{4,4}$ and higher order nonlinear Green's functions are identical
in the asymptotic regime to one of the two integrals appearing in the
3-order quantity.  The only differences are in the combinatorial
factors that account for how many ways we can choose the rungs to
connect between $p$ struts.

If the second rung is connecting the same struts as the rung before it
we have the same combinatorial factor as in the 1-loop order, namely
$p(p-1)/2$. This provides a contribution to $\zeta_p$ which is
$p(p-1)\tilde \delta^2 b_1/2$. If the second rung is {\em not}
connecting the same struts as the rung before it, we have
$p(p-1)(p-2)$ contributions. This is due to the existence of $p(p-1)/2$
ways connect two struts with the first rung, and then $2(p-2)$ ways to
connect one of these two struts with the remaining $(p-2)$ struts.
This leads to a contribution $p(p-1)(p-2)b_2\tilde \delta^2$.  We
should stress that the loops must have a joint strut to give a $\ln$
contribution. Two disconnected loops lead only to $\ln^2$
contributions, which do not enter the 2-loop corrections to the
scaling exponents.  In total we find
\begin{eqnarray}
\zeta_p&=&\frac{n}{3}+p \delta_a
-\frac{p(p-1)}{2}[\tilde \delta+b_1\tilde \delta^2]
\nonumber\\&-&p(p-1)(p-2)b_2\tilde
 \delta^2+O(\tilde \delta^3) \ . \label{zpaf}
\end{eqnarray}
Substituting Eqs.~(\ref{imp1}, \ref{imp2})
 we obtain finally
\begin{equation}
\zeta_p=\frac{p}{3}
-\frac{p(p-3)}{2}\delta_2[1+2\delta_2b_2(p-2)]+O(\delta_2^3)
\label{finalzn}
\end{equation}
We should stress that the functional form presented in this equation
is solid. It is shown in Appendix \ref{6rung} that the contribution
coming from 6-point irreducible rungs is only renormalizing the value
of $b_2$ which anyway depends on the precise analytic form of the 
4-point rung which is not available at the present time. 
We estimate the range of validity of this order of the calculation
by the H\"older inequalities, which disallow a nonlinear increase
in the $\zeta_p$ as a function of $p$. The inflection point where this
requirement is violated may serve as a good estimate for the range 
of validity.  This inflection point occurs at $p\approx 1.4-1/(6\, b_2\,
\d_2)\,\approx 12$. In Fig.~\ref{fig-5} we show, within this range, the K41
prediction, the 1-loop approximation (equivalent to K62) and our
2-loop final result. It is obvious that the 2-loop loop prediction
goes considerably beyond the range of validity of the K62 formula
which has an unphysical maximum at $p\approx 11.5$. We believe that
all the reliably measured values of $\zeta_p$ agree very well with
this prediction.

Using the bridge relation \cite{Fri} $\mu=2-\zeta_6$ we predict
\begin{equation}
\mu=9\delta_2(1+8b_2\delta_2) \ .
\end{equation}
Plugging in the numbers we get $\mu= 0.235+O(\delta_2^3)$. This is to
be contrasted with the K62 prediction $\mu\approx 0.27$. We conclude
that the 2-loop contribution is very significant for experimentally
measured exponents. If one wishes to obtain theoretical results for
$\zeta_p$ with higher values of $p$ one needs to consider the 3-loop
contributions, which pose no further conceptual
difficulties. Nevertheless the experimental situation does not warrant
at the present time the effort needed to accomplish such a
calculation.

One should stress before closing this Section that the form
of Eq.~(\ref{finalzn}) is universal, stemming from the structure
of the ladder diagrams and from combinatorics only. However the
numerical value of $b_2$ is model dependent. We have checked
that changing the form of the rung keeping the asymptotics
unchanged results in $b_2$ remaining negative while
its value not changing by more than a factor of 2 or so.
At this moment in time one can determine $b_2$ using the value of
$\zeta_4$ from experiments, allowing us then to predict
accurate values of $\zeta_n$ for $n$ up to 12. It is our
plan however to develop in the near future a theoretical
equations for the 4-point and 6-point rungs, leading to an ab-inito determination
of their analytic forms, and with them of the parameters $\tilde\delta$
and $b_2$.
\section{Summary and discussion}
\label{s:discuss}
The main steps of this and previous papers leading to the
present results have been as follows:
\begin{itemize}
\item The theory is developed using BL-velocities to eliminate the
spurious infrared divergences that are due to sweeping effects when
Eulerian velocities are employed.
\item The Dyson-Wyld perturbation theory was line resummed in order
to achieve order by order convergent perturbation theory with K41
propagators as the lines in the theory. At this point the objects of
the theory are two 2-point propagators (Green's function and
correlator) and one 3-point vertex. The 3-point vertex is in no way
``small", and renormalizing it does not change this fact \cite{93LL}.
\item Multipoint correlation functions are considered when
$p$ coordinates coalesce together. In the fusion limit $\kappa r \to
0$ it is advantageous to reorganize the theory in terms of one
propagator (K41 Green's function), and 4-point, 6-point
vertices etc. (the rungs). The series of diagrams contributing to the
fusion limit are then simple ladder diagrams.
\item The crucial step of the theory is achieved by two requirements: (i)
the 4-point rung should be consistent at the level of the {\em
skeleton diagrams} with the fusion rules with K41 scaling
exponents. (ii) The resummation of the ladder diagrams that appear
when 2 coordinates fuse together should lead to the {\em correct}
value of $\zeta_2$. These double requirements accomplish two things in
one go: (i) the theory is now developed {\em around the K41 limit},
leading to the appearance of the small parameter $\delta_2$ 
in front of the 4-point rung, and
(ii) all the anomalies are coming from the ladder resummations. The 6-point
rung is shown explicitly (Appendix~\ref{6rung}) to be of second order
in the small parameter, 8-point rungs are of third order, etc.
\item We computed the anomalous exponents in 1-loop order, inputting the
value of $\zeta_2$ and requiring that $\zeta_3=1$. The result is that the
scaling exponents are predicted to this order to agree with the log
normal model K62. We showed that to this order the result is universal,
independent of the simplifications and of the model form of the rung.
\item We computed the anomalous exponents in 2-loop order. The malaise
of K62 is cured, the 2-loop contribution has a sign that lifts up the
exponents from the down curve of the K62 parabola. While the form of
the 2-loop result is universal, the numerical value of the parameter
$b_2$ appearing in the final result is model dependent, with
contributions for the 4-point and 6-point rungs.
\end{itemize}

To improve upon the present theory one needs to develop a theory for
the 4-point and 6-point interaction amplitudes. Here we determined only the
asymptotic properties of the 4-point rung, and this allowed us to
predict the form of the scaling exponents, but an input of the value
of the anomalous part of $\zeta_2$ was needed to nail the 1-loop
order. In fact we could use the value of $\zeta_4$ to fix the value of
$b_2$ and gain a solid prediction of all the exponents to 2-loop
order. Such a prediction for $\zeta_n$ would be valid up to $n\approx
12$. It is very easy to generalize the result that we have to 3-loop
order, with the introduction of yet one more parameter associated with
the 3-loop integrals, say $b_3$, which included also contributions
from the irreducible 8-point rung. The result would read
\begin{eqnarray}
\zeta_n&=&\frac{n}{3} -\frac{n(n-3)}{2}\delta_2[1+2\delta_2(n-2)b_2
\nonumber\\&+&6\delta^2_2b_3(n-1)(n-2)] +O(\delta_2^4) \ .
\label{3loopzn}
\end{eqnarray}
We stress that this form stems from the structure of the ladder
diagrams, and we consider it very solid. From one point of view we can
now use the value of $\zeta_5$ to fix $b_3$ to provide a prediction
that is valid for any $n$ within experimental reach for quite some
time. But this is not the main point.  The main point is that we have
identified the coefficients appearing in this formula with particular
objects, i.e the 4-point and higher order vertices which appear in the
theory as the rungs of the ladders.  Obviously, a calculation of the
renormalized rungs from first principle would remove the need to input
experimental information altogether, affording us a complete theory of
the scaling exponents of isotropic turbulence. At this point this is
still not in the cards.
\acknowledgments
It is a pleasure to thank Anna Pomyalov for her patient help with the
diagrams in this paper.  We thank her, Yoram Cohen, Ayse Erzan and
Massimo Vergassola for useful comments on the manuscript. This work
has been supported in part by the Israel Science Foundation, the
German-Isreali Foundation, the European Commission under contract
HPRN-CT-2000-00162 (``Nonideal Turbulence"), and the Naftali and Anna
Backenroth-Bronicki Fund for Research in Chaos and Complexity.
\appendix
\section{Explanation of the diagrammatic expansion in Figs.~2, 3}
\label{ap-A}
It is important to stress that in the present theory we take into
account all the necessary contributions. To understand this we need
to say a few more words about the representation of $G_{2,2}$ in terms
of simple ladders only, and of the $n$th order correlation function
with $p$ fused coordinates as shown in Fig.~\ref{fig-2}.

The natural objects in the straightforward perturbation theory are the
2-point correlation function and Green's function (\ref{b6}), and the
3-point vertex resulting from the Navier-Stokes nonlinearity $\B.u\cdot
\B.\nabla \B.u$. After line resummation the theory contains ``dressed"
correlator and Green's function. The 3-point vertex is protected by
Galilean invariance and is not effected much by dressing
\cite{93LL}. Thus, when we write the expansion of $G_{2,2}$ many
diagrams involving these objects appear. The strategy that leads to
the simple ladder expansion of Fig.~\ref{fig-1} is as follows: Every
diagram that contributes to the series is inspected for its cross
section, or in other words what are the kind of objects that intersect
a line cutting across the diagram.  The line is put at the
left of the diagram, and is moved to the right.  Every time that the
line intersects two Green's functions that are oriented as shown in
Fig.~\ref{fig-1} we mark that position, and move the line 
further to the right, until we intersect again two Green's functions,
etc. For every pair of such intersections we now sum up all the
topologically allowed diagrams that can be inflated ad infinitum from
the fragments appearing between the two pairs of Green's
functions. This infinite resummation is the representation of the
rung, which is actually a 4-point vertex. This procedure is flawless,
taking into account all the possible diagram in the series of
$G_{2,2}$ {\em except} for one subseries. This is the subseries of
diagrams in which the cross section contains exactly two 2-point
correlation functions.  It was shown in Paper II that this can happen
only {\em once per diagram}, and therefore we cannot resum such
contributions to the rung, since this will lead to one rung differing
from all the others. Thus the series shown in Fig.~\ref{fig-1}a
contains in the $+\dots$ also ladders in which the struts contain two
correlators above each other. With these we account for all the
possible diagrams in $G_{2,2}$.

\begin{figure}
\epsfxsize=8.6truecm
\epsfbox{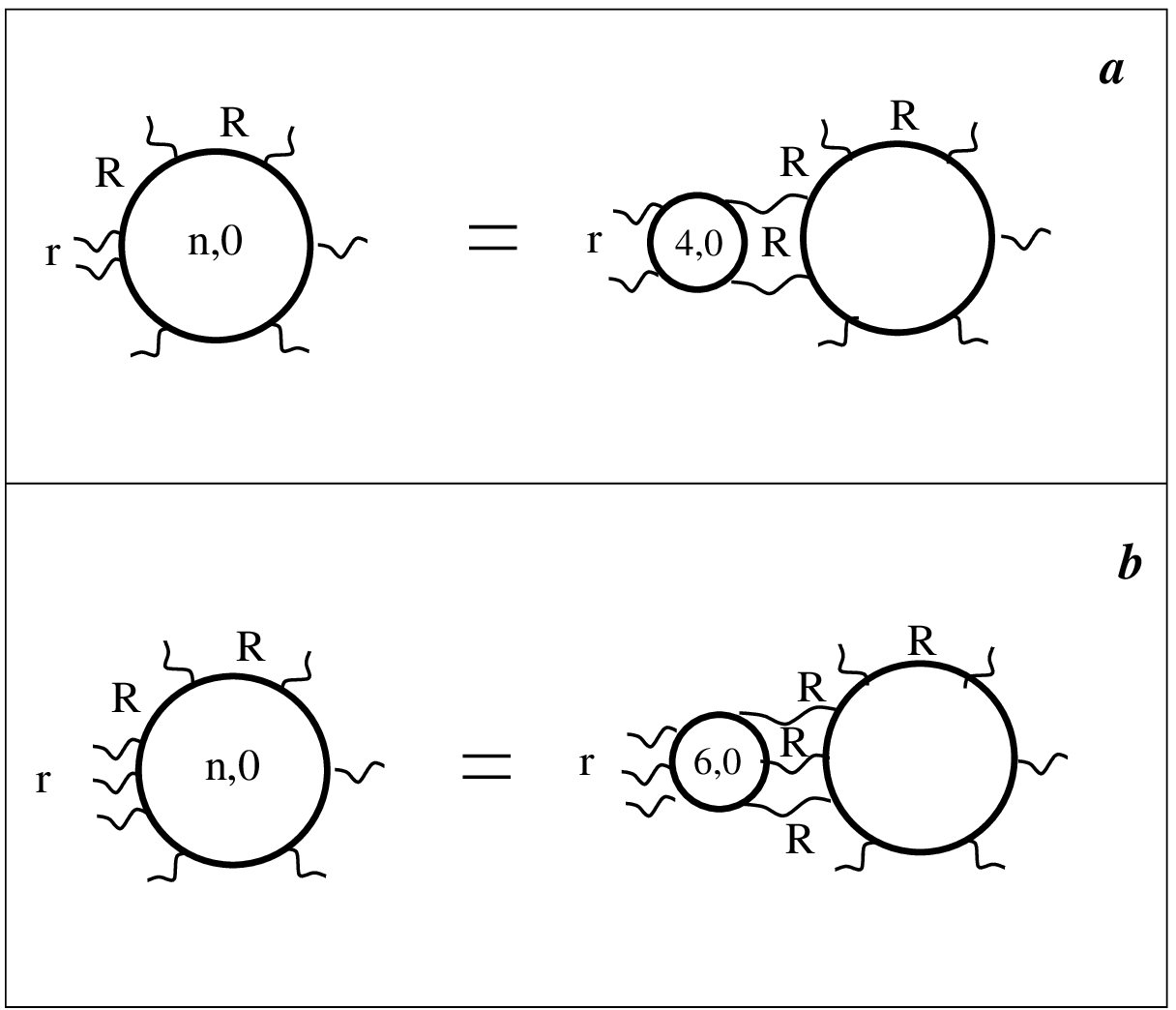}
\vskip 0.3 cm
\caption{The contributions to the fusion of 2 and 3 coordinates
from diagrams containing two and three correlators in the
cross section.}
\label{fig-6}
\end{figure}

The presence of these diagrams also complicates the discussion of the
fusion rules and Fig.~\ref{fig-2}. When we pool out the fragment
containing the two coalescing coordinates we will generate in the
series expansion of this fragment also the diagrams containing two
correlators in the cross section. To understand their roles we will
always consider the fragment of the diagrams to the right of this
special cross section as belonging to the main body, see
Fig.~\ref{fig-6}. But now the fragment pulled out becomes a 4th order
correlation function instead of $G_{2,2}$.  One can analyze the role
of this diagram in the fusion limit, and the conclusion is that when
$r\ll R$ it contributes the same power $(R/r)^{\zeta_2}$ as the
diagrams considered in the body of the paper. We thus conclude that
the procedure followed in the body of the paper is amply sufficient
for the calculation of $\zeta_2$.

The same type of considerations apply when we fuse $p$ coordinates.
In the series expansion for $G_{p,p}$ we will have however ladder
diagrams whose topology requires resummation into 2$p$-point
irreducible interaction amplitudes which serve as new types of
rungs. The first one appears at the level of $G_{3,3}$, and is the
6-point rung that is discussed explicitly in Appendix~\ref{6rung}. It
will be argued that the 6-point rung is of second order in the
smallness $\delta_2$ that characterizes the 4-point rung. Respectively
the 2$p$-point rungs are of $O(\delta_2^{p-1})$ and thus rungs with
$p>3$ will not affect the analysis of this paper.

On top of the diagrams considered in the body of the paper one needs
to consider those having $p$ correlators in the cross section. The
conclusion is however the same - the diagrams considered give all the
necessary information, no new (or more divergent) information is
available in the diagrams that we do not consider explicitly.

At the end of the day the procedure of calculating the scaling
exponents to $O(\delta_2^2)$ leaves us with only {\em three} dressed
objects in the theory, the dressed 2-point Greens' function and new
4-point and 6-point dressed vertices which we call the rungs. We will
show that the rungs are {\em small} and therefore these three objects
suffice for a consistent and controlled theory of the scaling
exponents.

A reader who is an expert in diagrammatic theories may feel worried
that the procedure advocated here mixes into the 4-point vertex some
2-particle reducible diagrams. This is indeed so.  In other words (and
see Paper III for more details) there are contributions in which the
cross section includes two propagators, like one Green's function and
one correlator, or two Green's functions oriented oppositely to the
ones plotted in Figs.~\ref{fig-1} and \ref{fig-2}, which we resum into
our 4-point vertex. Usually one would prefer not to include 2-particle
reducible diagrams in the 4-point vertex, but rather to distinguish
different 4-point vertices. In our case one would need three different
4-point rungs, all presented in detail in Paper III. Indeed, if we
planned to compute our 4-point vertices from summing up diagrams it
would be very advisable to distinguish different types of rungs.  In
this paper we determine however the properties of the 4-point vertex
from the fusion rules, and it makes no difference at all how we
classify the diagrams. For the sake of clear presentation it is much
better to have just one type of rung. This rung is 2-particle
reducible and hides three types of 2-particle irreducible rungs as a
single object. The conclusions of the analysis are independent of this
simplification.
\section{Resummation into diagonal K41 propagators}
\label{K41prop}
The starting point of this rearrangement are the mass operators in
$k,\omega$ representation
$\Sigma_{\alpha\beta}(\B.r_0|\B.k_1,\B.k_2,\omega)$ and
$\Phi_{\alpha\beta}(\B.r_0|\B.k_1,\B.k_2,\omega)$. Define the
``diagonal" part of the mass operators as
\begin{eqnarray}
\sigma_{\alpha\beta}({\B.k_1+\B.k_2\over 2})&\equiv&\int
{d(\B.k_1-\B.k_2)\over (2\pi)^3}
\Sigma_{\alpha\beta}(\B.k_1,\B.k_2,0) \ , \\
\phi_{\alpha\beta}({\B.k_1+\B.k_2\over 2})&\equiv&\int
{d(\B.k_1-\B.k_2)\over (2\pi)^3}
\Phi_{\alpha\beta}(\B.k_1,\B.k_2,0) \ .
\end{eqnarray}
In these definitions $\B.r_0$ disappears. The reason is that for
objects which are time independent the Eulerian and BL-representations
are equivalent and the designation $\B.r_0$ is unneeded.  Here we have
objects with $\omega=0$, or time-integrated quantities.  It was shown
in ref.~\cite{demonstration} that time integrated quantities are
related to simultaneous correlations, and as such they lose the
$\B.r_0$ designation.

Denote the rest of the mass operators as
\begin{eqnarray}\nn 
\tilde\Sigma_{\alpha\beta}(\B.r_0|\B.k_1,\B.k_2,\omega)
&\equiv&  \Sigma_{\alpha\beta}(\B.r_0|\B.k_1,\B.k_2,\omega)
\sigma_{\alpha\beta}({\B.k_1+\B.k_2\over 2})\ , \label{defsig} \\ \nn
\tilde\Phi_{\alpha\beta}(\B.r_0|\B.k_1,\B.k_2,\omega)
&\equiv&   \Phi_{\alpha\beta}(\B.r_0|\B.k_1,\B.k_2,\omega) -
\phi_{\alpha\beta}({\B.k_1+\B.k_2\over 2})\ .
\end{eqnarray}
For translationally invariant tensors in homogeneous and
incompressible turbulence one can write:
\begin{eqnarray}
&&\sigma_{\alpha\beta}(\B.k) =P_{\alpha\beta}(\B.k) \sigma(k)\ , \\ \nn
&&\phi_{\alpha\beta}(\B.k) =P_{\alpha\beta}(\B.k) \phi(k) \ ,
\end{eqnarray}
where $P_{\alpha\beta}(\B.k)$ is the transverse projector,
\begin{equation}
P_{\alpha\beta}(\B.k) =\delta_{\alpha\beta}
-\frac{k_\alpha k_\beta}{k^2}\ . \label{Pink}
\end{equation}
It is known \cite{61Wyl} that $\sigma(k)$ (which is the
mass operator taken at $\omega=0$) is purely imaginary
\begin{equation}
\sigma(k) = -i\gamma(k) \ ,
\end{equation}
with $\gamma(k)$ real positive. On the other hand $\phi(k)$ is purely
real.  The diagrammatic series expansion of both $\gamma(k)$ and
$\phi(k)$ converge order by order, and using scaling relations as
shown in (\ref{b11}) one can find their scaling behavior. The
order-by-order theory dictates a K41 evaluation of these objects which is
\begin{eqnarray}
\gamma(k)&=&c_\gamma [\bar\epsilon k]^{2/3}\ , \\
\phi(k)&=&c_\phi\bar\epsilon k^{-3} \ ,
\end{eqnarray}
where $c_\gamma$ and $c_\phi$ are dimensionless constants.

The Dyson-Wyld equations can be written shortly as
\begin{eqnarray}
&&(\omega+i\nu k^2) \B.G =\B.P+ \B.\Sigma*\B.G\ , \label{dysonshort}\\
&&\B.F=\B.G*(\B.\Phi+\B.D)*\B.G \ , \label{wyldshort}
\end{eqnarray}
where $\nu$ is the molecular viscosity, $\B.P$ is the transverse
projector, and $\B.D$ is the correlation function of the external
force which is localized in the energy containing interval. The symbol
$*$ stands for summation over tensor indices and integration over
intermediate $\B.k$. Substituting $\B.\Sigma$ from Eq.~(\ref{defsig})
into the Dyson equation we rewrite:
\begin{equation}
[\omega+i\nu k^2+i\gamma(k)] \B.G =\B.P+ \tilde\B.\Sigma*\B.G \ .
\end{equation}
In the bulk of the inertial interval we can neglect $\nu k^2$ with
impunity.  The zero order solution of this equation is obtained by
neglecting $\tilde\B.\Sigma$:
\begin{eqnarray}
G_{\alpha\beta}\to g_{\alpha\beta}(\B.k,\omega)
&=&P_{\alpha\beta}(\B.k) g(k,\omega)\ , \\
g(k,\omega)&=&\frac{1}{\omega+i\gamma(k)}\ .
\end{eqnarray}
The zero order solution of $\B.F$ is obtained in three steps: first
replace $\B.\Phi$ by $\B.\phi$, secondly neglect $\B.D$ in the
inertial interval in comparison with $\B.\phi$, and lastly substitute
$\B.g$ instead of $\B.G$ in Eq.~(\ref{wyldshort}). The result is
\begin{eqnarray}
F_{\alpha\beta}\to f_{\alpha\beta}(\B.k,\omega)&=&P_{\alpha\beta}(\B.k)
f(k,\omega) \ , \\
f(k,\omega)&=&\frac{\phi(k)}{\omega^2+\gamma^2(k)} \ .
\end{eqnarray}

Iterating Eqs.~(\ref{dysonshort}, \ref{wyldshort}) without the bare
forcing and viscosity results in a new diagrammatic series, which
topologically is exactly the same as the old Wyld diagrammatic
expansion before line re-summation. The difference is twofold. First,
instead of bare propagators we have K41 propagators $\B.g$ and $\B.f$,
and every 1-particle reducible fragment of any diagram will have a
counter term which subtracts its ``diagonal" part. This counter term
is of no consequence for our procedure here since the diagrams
involving it are resummed in the 4-point vertices (the rungs) together
with all the other contributions as explained in Appendix~\ref{ap-A}.
The resulting topological structure of the ladder diagrams is thus
unchanged in the new formulation.
\section{Self consistency at the level of K41}
\label{ap-self}
Before establishing this self consistency we need to pass from
correlation functions in $\B.k,\omega$ representation to structure
functions. The theory is done naturally in $\B.k,\omega$
representation but the experimental scaling exponents are measured in
simultaneous structure functions. We first transform from
$\o$-representation of $p^{\rm th}$-order correlation function
$\BC.F_p(\{\B.k_j,\o_j\})$ to simultaneous correlation function
$\B.F_p(\{\B.k_j\})$ by the integration:
\BE.omega-t
\B.F_p(\{\B.k_j\})=\int\limits_{-\infty}^\infty\prod_{i=1}^p
\frac{d\o_i}{2\pi}\d(\o_1+\dots \o_p)\BC.F_p(\{\B.k_j,\o_j\})\ .
\EE
Here $\{\B.k_j,\o_j\}$ and $\{\B.k_j\}$ are sets of corresponding
variables with $j=1, \dots p$. The transformation from $\B.k$
representation of $\B.F_p(\{\B.k_j\})$ to the $p^{\rm th}$-order
structure function is done as follows: define the longitudinal
component of the velocity as
\begin{equation}
S_p(r)=\Big\langle \Big\{\big[\B.u(\frac{\B.r}{2})
-\B.u(-\frac{\B.r}{2})\big]\cdot
\frac{\B.r}{r}\Big\}^p\Big\rangle \ .
\end{equation}
Each of the factors is Fourier transformed according to
\begin{eqnarray}
&&[\B.u(\frac{\B.r}{2})-\B.u(-\frac{\B.r}{2})\big]
=\int \frac{d\B.k_j}{(2\pi)^3} \hat\B.u
(\B.k_j)\nonumber\\&\times&\Big[
\exp\big(i\frac{\B.k_j\cdot \B.r}{2}\big)
-\exp\big(-i\frac{\B.k_j\cdot \B.r}{2}\big)\Big]\,,
\end{eqnarray}
Accordingly,
\begin{eqnarray}
\label{fdiff}\\
\label{k-r}
S_p(r) &=&(2\pi)^3\!\int\limits_{-\infty}^\infty\prod_{i=1}^p
\frac{d\B.k_i}{(2\pi)^3}\d(\B.k_1
 \!+\dots \B.k_p)\nonumber\\ &\times& f_p(\B.r,\{\B.k_j\})
F_p(\{\B.k_j\})\ .
\end{eqnarray}
Here
\begin{equation}
(2\pi)^3F_p(\{\B.k_j\})\d(\B.k_1 \!+\dots \B.k_p)
=\Big\langle\prod _{j=1}^p
\hat\B.u(\B.k_j)\cdot \frac{\B.r}{r}\Big\rangle \ .
\end{equation}
The functions $f_p(\B.r,\{\B.k_j\})$ are seen
 from Eq.~(\ref{fdiff}) to be:
\BE.str
f_p(\B.r,\{\B.k_j\})=\prod_{j=1}^p  [2i  \sin
(\case{1}{2}\B.k_j\!\cdot \! \B.r)] \ .
\EE
In the limit $\B.r  \to 0$
 \BE.lim
f_p(\B.r,\{\B.k_j\})\propto  \prod_{j=1}^p
 (\B.k_j\!\cdot \! \B.r)\ .
\EE
The K41 scaling exponents $y_p$ associated with
 $p^{\rm th}$-order correlation function
 $\BC.F_p(\{\B.k_j,\o_j\})\propto k^{-y_p}$ in
 $(\B.k,\omega)$-representation is
\BE.expy
y_p=4p-11/3\ .
\EE
 This corresponds to $\B.S_p(\B.r)\propto r^{p/3}$ {\em under the
 condition of convergence of integrals} (\ref{omega-t}, \ref{k-r}).

Next consider the 3rd order Green's function,
$G_{3,3}(\{k_j,\kappa_j\})$ in which we denoted by $k_j$ the set of
incoming wave vectors and by $\k_j$ the set of outgoing wave vectors.
The skeleton diagram of $G^{\rm s}_{3,3}(\{k_j,\kappa_j\})$ which
involves 4-point rungs is shown as diagram (3) in
Fig.~\ref{fig-3}a. (The contribution of 6-point rungs to the skeleton
is considered in Appendix~\ref{6rung} and shown not to change the
present considerations).  This skeleton has two rungs, and we consider
it in the limit that the incoming $k_j$ vectors are much larger than
the outgoing $\k_j$. In this limit we have four Green's functions with
large $k$, contributing $\gamma_k^{-4}$, and one vertex with all $k$
large, contributing $k$. The two rungs have large $k$ vector in them
[$k_5$ in Eq.~(\ref{rung})], giving $k^6$. Finally, one of the rungs
has large $k$ coming and going, and Eq.~(\ref{rung}) requires for it a
$k^{2/3}$. Altogether this gives $G_{3,3}(\{k_j,\kappa_j\})\propto
k^{x_3}$ with $x_3=25/3$ which is equal to $y_3$ given by
Eq.~(\ref{expy}). This means that the skeleton diagrams for
$G_{3,3}(\{k_j,\kappa_j\})$ (with asymptotics of the rung defined by
the two-point fusion rules) automatically reproduces the K41 scaling
exponent $\zeta_3=1$ in the three-point fusion. This is true {\em
subject to the condition} that the integrals (\ref{omega-t},
\ref{k-r}) for $p=3$ converge. That this is so may be shown by a
direct calculation. For future purposes it is extremely important to
note that the principal contribution to the $\B.k$ integral
(\ref{k-r}) comes from the region where $k_1\sim k_2\sim k_ 3 \sim
1/r$.

Now let us compare diagram (3) in Fig.~\ref{fig-3}a and
Fig.~\ref{fig-4} with the skeleton diagrams for $G_{3,3}$ and
$G_{4,4}$.  One recognizes that in general for $G_{p,p}$, we will have
$(p-1)$ rungs with large incoming $k$, contributing $k^{-3(p-1)}$
[originating from $k_e$ in Eq.~(\ref{rung})].  We will have also
$2p-2$ Green's functions with large $k$ contributing $k^{-(2p-2)2/3}$.
Next we will have $p-2$ outgoing legs with large $k$ contributing
$k^{-(p-2)2/3}$ from Eq.~(9). Finally we will have $2p-2$ vertices
having incoming and outgoing large $k$ vectors, contributing
$k^{p-2}$. All together we find that $G_{p,p}(\{k_j, \kappa_j\})
\propto k^{-x_p}$ with $x_p=4p-11/3$ which is equal to $y_p$ given by
Eq.~(\ref{expy}). Convergence of the $\B.k$ integral (\ref{k-r}) for
$p=4$ may be shown by direct calculations. A proof of convergence of
the $\B.k$ integrals (\ref{k-r}) for $p>4$ is a tedious exercise which
nevertheless may be done, for example, iteratively. It is readily
demonstrated that the integral converges when all $k_j$-vectors are of
the same order of magnitude (say, $k$). Then $G_{p,p}\sim k^{11/3-
4p}$. After $(p-1)$\ $\o$-integrations (each of them giving a factor
$k^{2/3}$) one has $k^{3(p-1)-p/3}$ which is enough for convergence of
$(p-1)$ $d^3 k$ integrals in the UV region $k_j\sim k \gg 1/r$.  In
the IR region $k_j\sim k \ll 1/r$ the functions $f_p$ provide the
integral with additional $k^p$ factor [according to (\ref{lim})] which
guarantees the convergence.

The considerations of the 6-point and higher order rungs leave these
conclusions invariant.
\section{Analysis of two loop integrals contributing to $\zeta_2$}
\label{2lint}
The integrand in the integral (\ref{Kkp}) is a function of $q_1$ and
$q_2$ and it depends on $k$ and $\kappa$ as parameters. The
integration range is the $q_1-q_2$ infinite plane, but in the limit
$k\gg \kappa$ the main contribution comes form the four finite
quadrants $\kappa< |q_1|,|q_2|<k$. Well inside the quadrants we are
allowed to use the asymptotic form in which $\kappa\ll |q_1|,|q_2|\ll
k$.  In this regime the integrand is $k,\kappa$-independent, and the
dependence of the integrals on $k,\kappa$ appears only via the limits
of integration. By changing the dummy variables $q_1$ and $q_2$ we can
now project all four quadrants into one of them, say $q_1$ and $q_2$
positive. In this asymptotic regime we can use for the rungs in the
integrand of (\ref{Kkp}) that include either $k$ or $\kappa$ their
asymptotic form (\ref{rung6}). This results in
\begin{eqnarray}
K(k,\kappa)&=&\tilde \delta^2\int_p^k\frac{dq_1}{q_1}
\int_p^k\frac{dq_2}{q_2} \Psi(q_1,q_2)\ ,
\label{2loop}
\\
\Psi(q_1,q_2)&=&\tilde \Psi(q_1,q_2)-\tilde \Psi(-q_1,q_2)\ .
\end{eqnarray}
In Appendix \ref{extract} we show how to analyze this kind of integral
with the aim of extracting the coefficients of the leading and first
subleading logarithmic terms, {\em i.e.}
\begin{equation}
K_1(k,\kappa) = \frac{a_1}{2}\ln^2 {(k/\kappa)} +b_1\ln {(k/\kappa)}
\end{equation}

 Using the results there (\ref{resulta}) with $\tilde
\Psi(q_1,q_2)=\tilde \Psi_1(q_1,q_2)$,
\begin{equation}
\tilde \Psi_1(q_1,q_2)=\frac{q_1^3|q_1- q_2|
{\rm sign}(q_2)}{2(q_1^2-
q_1q_2+q_2^2)^2} \ ,
\end{equation}
one find immediately $a\to a_1= 1$ as required by the anticipated
expansion employed in Eqs.~(\ref{2nd-2})-(\ref{2nd-sum}). To compute
$b_1$ we examine the integral $b_1(A)$ numerically, see
Fig.~\ref{fig-10}. We see that the requested limit exists and that
$b_1\approx -0.434$.
\begin{figure}
\epsfxsize=8.6truecm
\epsfbox{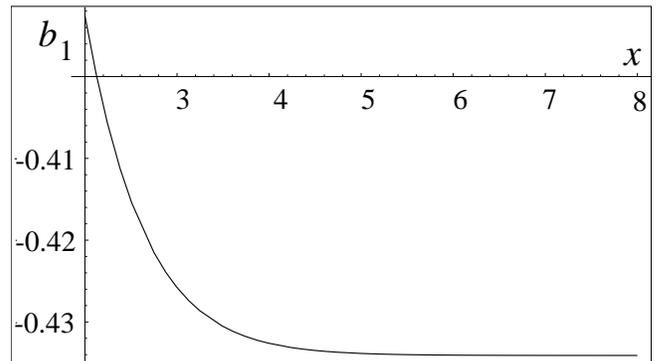}
\caption{Numerically computed dependence of $b_1(A)$ with $A=\exp(x)$.}
\label{fig-10}
\end{figure}

\section{Extraction of the subleading logarithmic term
from the 2-loop integrals}
\label{extract}
The 2-loop integrals have the characteristic structure appearing in
(\ref{2loop})
\begin{equation}
I(A) =\tilde\delta^2\int_1^A\frac{dq_1}
{q_1}\int_1^A \frac{dq_2}{q_2} \Psi(q_1,q_2) \ ,
\label{canonical}
\end{equation}
where $A\gg 1$ and $\Psi(q_1,q_2)$ is homogeneous function of degree
zero: $\Psi(\lambda q_1,\lambda q_2)=\Psi(q_1,q_2)$. When
$\Psi(q_1,q_2)=1$ then $I(A)=\ln^2{A}$. In general only the leading term of
$I(A)$ is proportional to $\ln^2{A}$ and we expect the following
subleading terms:
\begin{equation}
I(A)=\tilde \delta^2[\frac{a}{2}\ln^2{A}
+b\ln{A} +c +\frac{d}{A} +\dots ]\label{IA}
\end{equation}
Our goal is to find the coefficient $b$ in the limit $A\to
\infty$. Taking the first derivative of (\ref{IA})
 with respect to $A$
and multiplying by $A$ we find
\begin{eqnarray}
&&a\ln A+b-\frac{d}{A}-\dots=\int_1^A \frac{dq_2}{q_2} \Psi(1,q_2)+
\int_1^A \frac{dq_1}{q_1} \Psi(q_1,1)\nonumber\\
&&=\int_{1/A}^1\frac{dx}{x} \Psi(x,1)
+\int_{1/A}^1\frac{dy}{y} \Psi(1,y) \ ,
\label{ab}
\end{eqnarray}
where we changed the dummy variables $q_1=xA$ and $q_2=yA$. Taking
another derivative and multiplying by $A$ we find for large $A$
\begin{equation}
a=\Psi(1,\frac{1}{A})+\Psi(\frac{1}{A},1) \ . \label{resulta}
\end{equation}
Substituting this result in (\ref{ab}), and
representing $\ln{A}$ as $\int_{1/A}^1dx/x$
we find
\begin{eqnarray}
b&=&\lim_{A\to \infty} b(A) \ , \label{resultb}\\
b(A)&=&\int_{1/A}^1\frac{dx}{x}\Big[\Psi(x,1)+\Psi(1,x)-
\Psi(\frac{1}{A},1)-\Psi(1,\frac{1}{A})\Big] \ . \nonumber
\end{eqnarray}
If the expansion assumed in Eq.~(\ref{IA}) is valid, this limit must
exist.
\section{The nine 2-loop diagrams of $G_{3,3}$}
\label{ap-nine}

Consider diagram (1a) in Fig.~\ref{fig-7}. We are interested in the
ratio of $T_{3,1a}^{(2)}/T_3^{\rm s}$, where $T_{3,1a}^{(2)}$ is
obtained by substituting the diagram (1a) instead of $G_{3,3}$ in
Eq. (\ref{Tpdef}). In the asymptotic regime $\kappa r\ll 1$ the loop
integrals over $q_1$ and $q_2$ contribute mostly in the regime $k\gg
q_1,q_2\gg \kappa$. In this regime the integrals over
$k_a,\omega_a,k_b,\omega_b$ cancel in the desired ratio. Similarly the
Green's functions $G(\kappa_d)$, $G(\kappa_e)$ and $G(\kappa_f)$ also
cancel in the ratio.  Accordingly

\begin{figure}
\epsfxsize=8.6truecm
\epsfbox{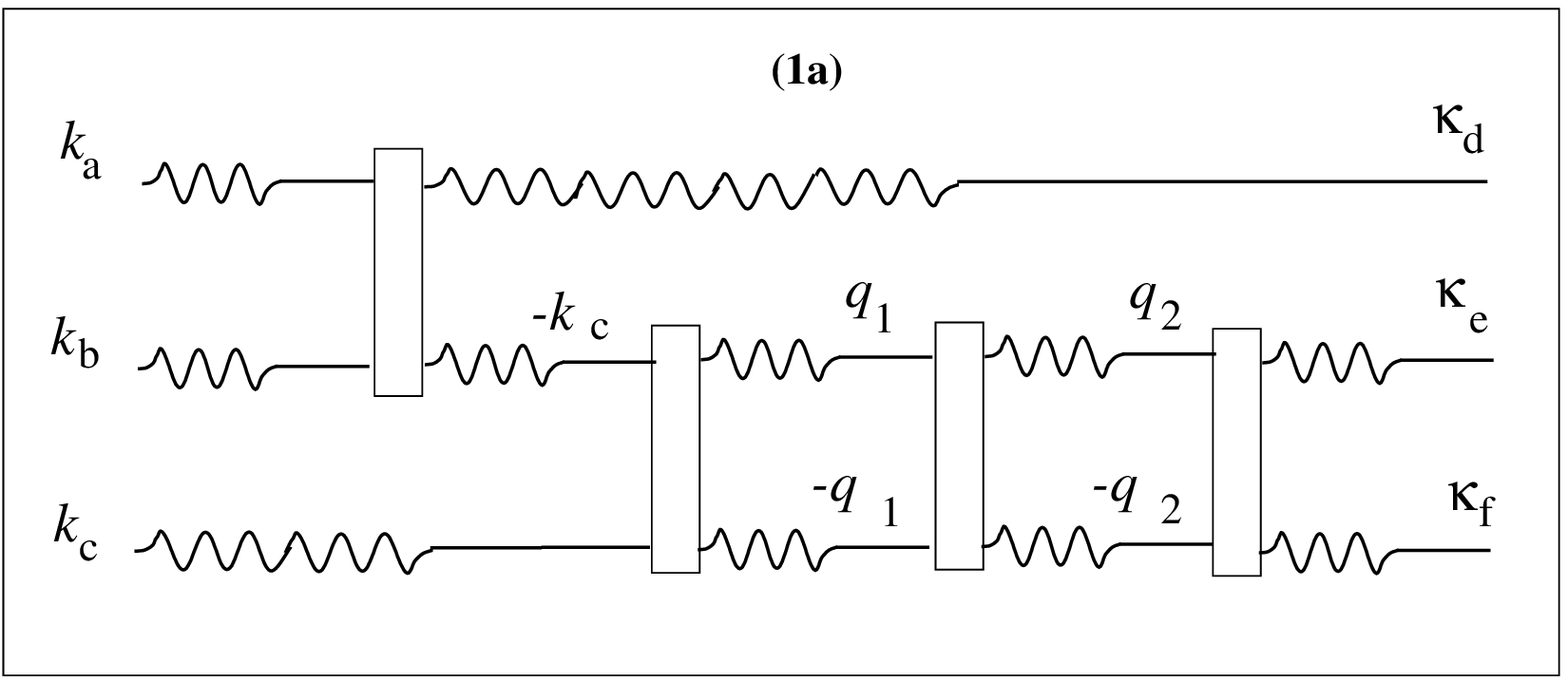}
\epsfxsize=8.6truecm
\epsfbox{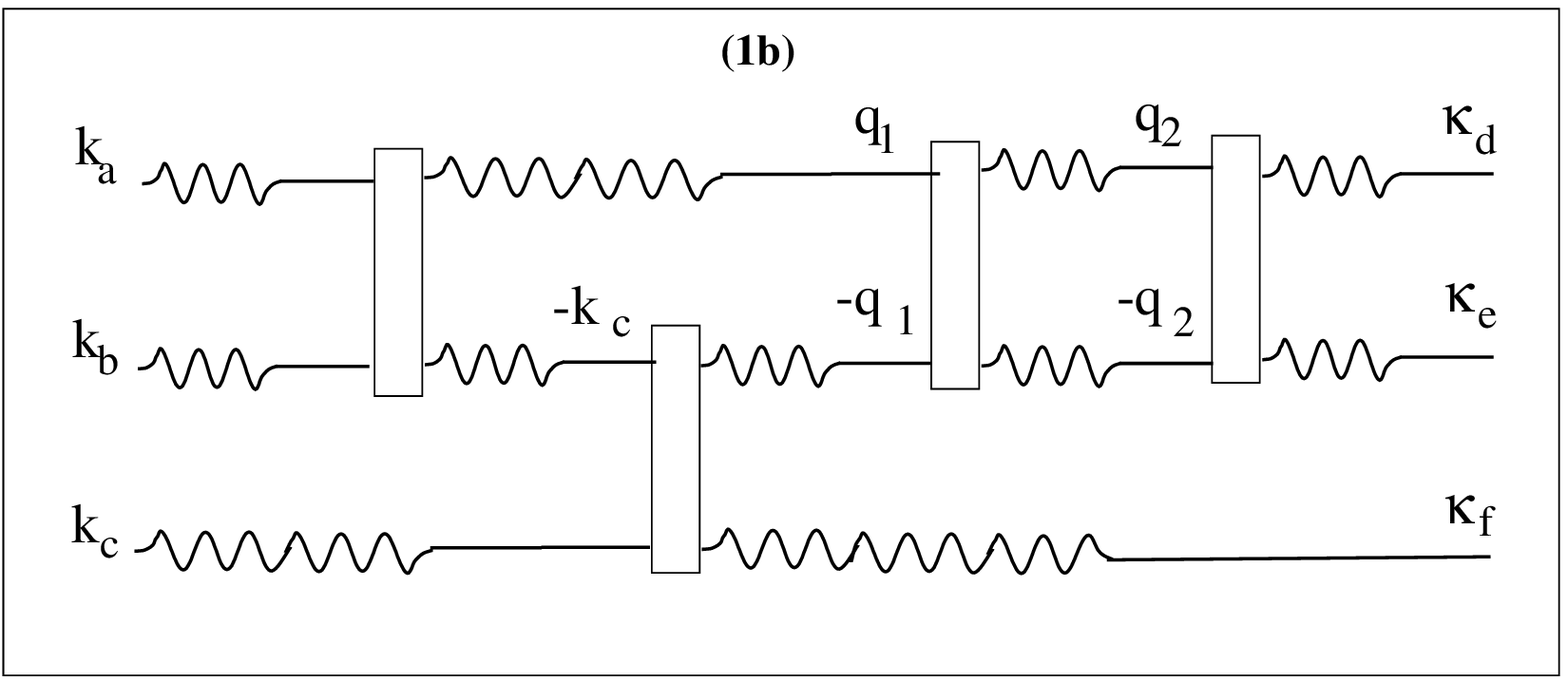}
\epsfxsize=8.6truecm
\epsfbox{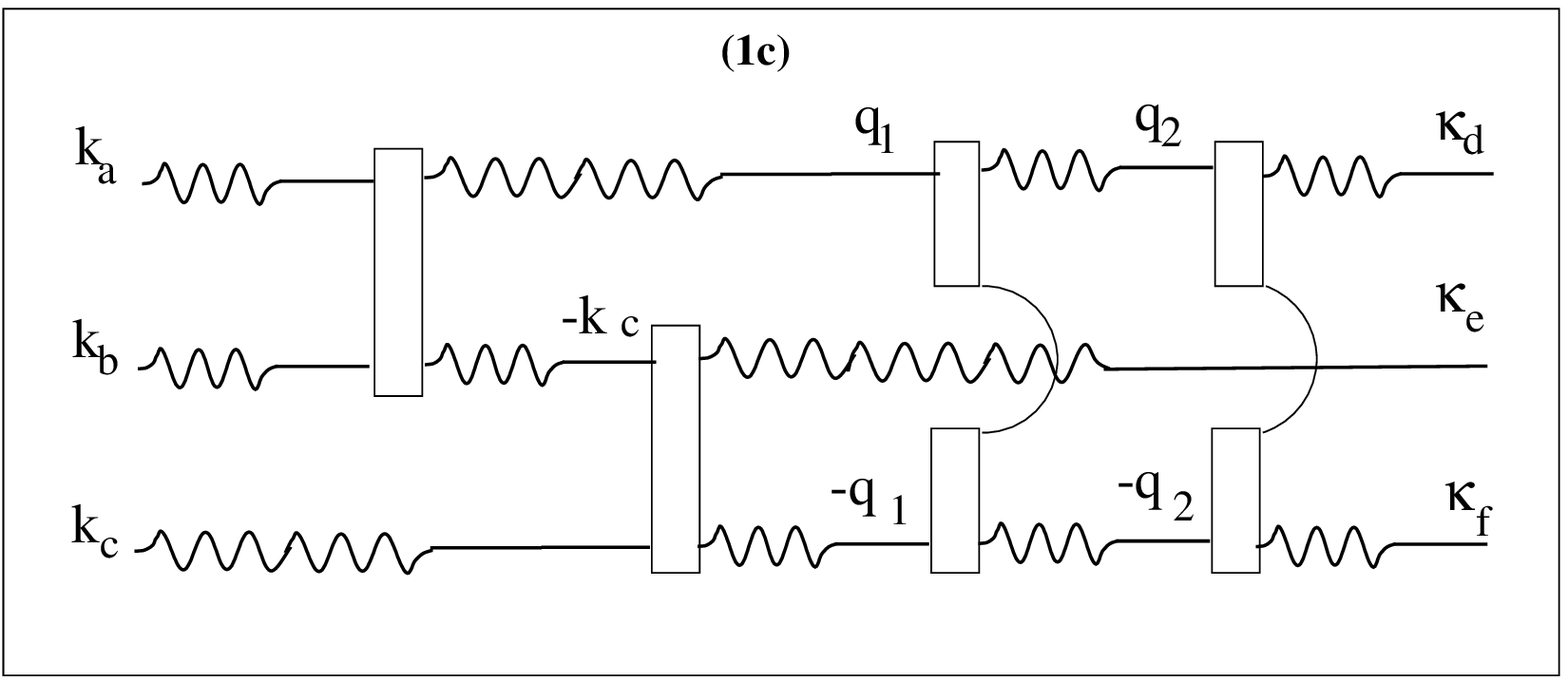}
\vskip 0.3 cm
\caption{The first group of three
2-loop diagrams appearing in the loop expansion of
$G_{3,3}$.}
\label{fig-7}
\end{figure}\noindent
$T_{3,1a}^{(2)}/T_3^{\rm s}$ can be
calculated from the {\em amputated} diagram (2) in Fig.~\ref{fig-9}, in
which the explicit dependence on $k_j$ and $\kappa_j$ has
disappeared. These wavevectors remain only in the limits of the
integrals over $q_1$ and $q_2$, with $k$ replaced by $1/r$. In this
diagram every black dot contributes a factor of $q_j^{1/3+\delta_a}$
where $q_j$ is the wavevector on the right of the black dot. This is a
remnant of the corresponding rung before the amputation. The thin line
connecting these dots is just a reminder that we have loop integrals
to perform. 

The point to understand now is that if we use diagrams (1b) and (1c)
in Fig.~\ref{fig-7} to form $T_{3,1b}^{(2)}$ and $T_{3,1c}^{(2)}$, the
ratio of these to $T_3^{\rm s}$ can be again calculated from the
amputation of their own diagrams.  This will lead to the {\em
identical} amputated diagram (2) of Fig.~\ref{fig-9}.  In addition,
and most importantly, the integral that needs to be computed is the
same as Eq.~(\ref{Kkp}). Thus one recaptures Eq.~(\ref{Kkp1}) but with
the combinatorial factor 3 in front of the RHS:
\begin{eqnarray}
T_{3,1a+1b+1c}^{(2)}(r,\kappa)&=&3\tilde
\delta^2\Big[\frac{1}{2}\ln^2{\Big(\frac{1}{\kappa r}\Big)}\nonumber \\
&+&b_1\ln{\Big(\frac{1}{\kappa
r}\Big)}\Big] T_3^{\rm s}(r,\kappa)
\ , \label{Kkp1a}
\end{eqnarray}
with $b_1$ of Eq.~(\ref{cb1}).

\begin{figure}
\epsfxsize=8.6truecm
\epsfbox{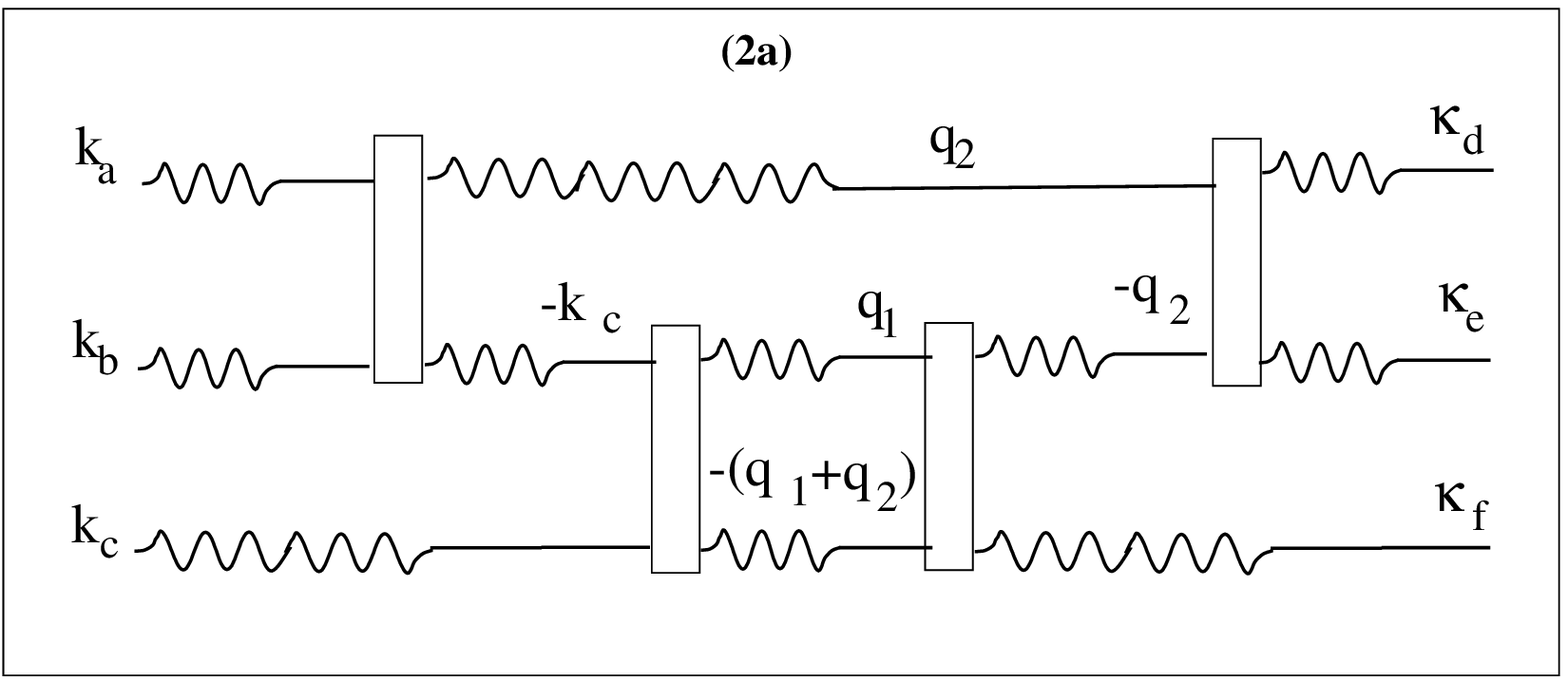}
\epsfxsize=8.6truecm
\epsfbox{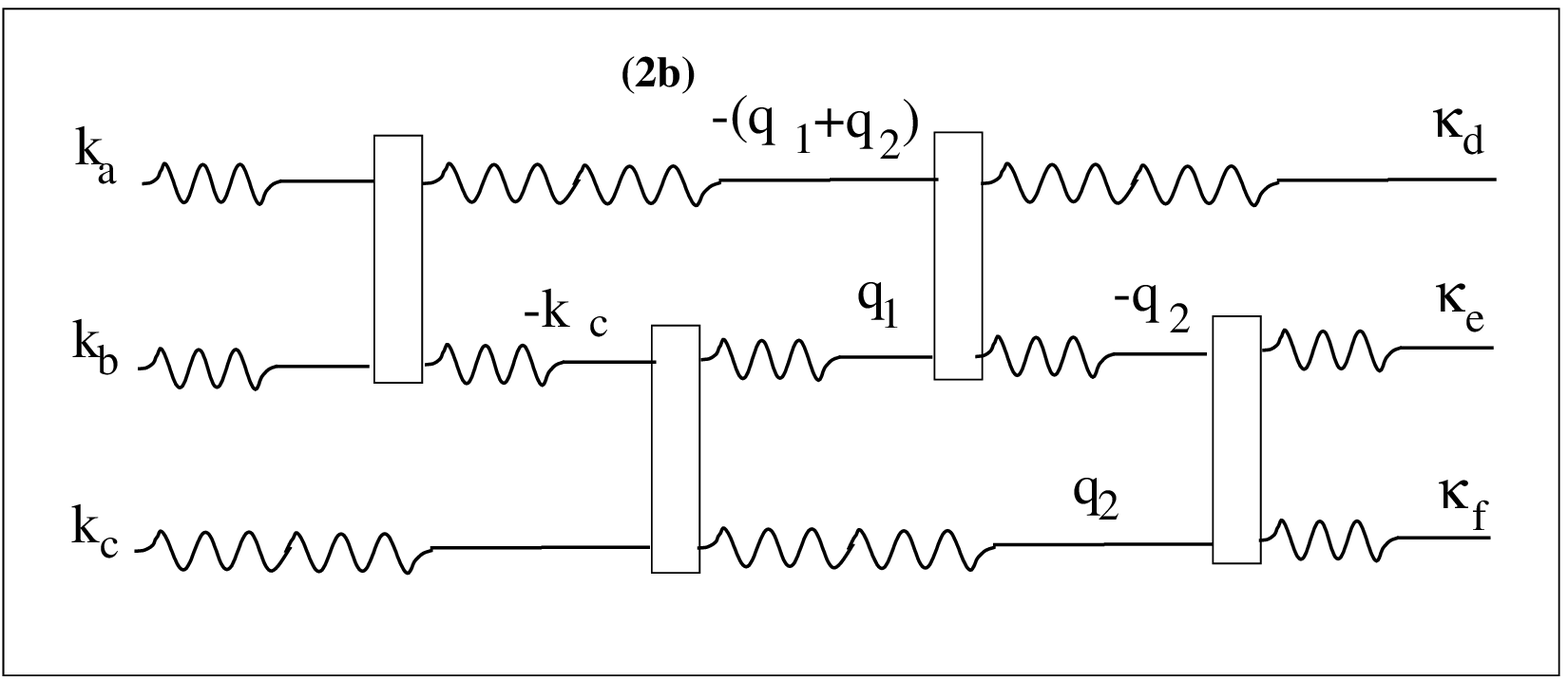}
\epsfxsize=8.6truecm
\epsfbox{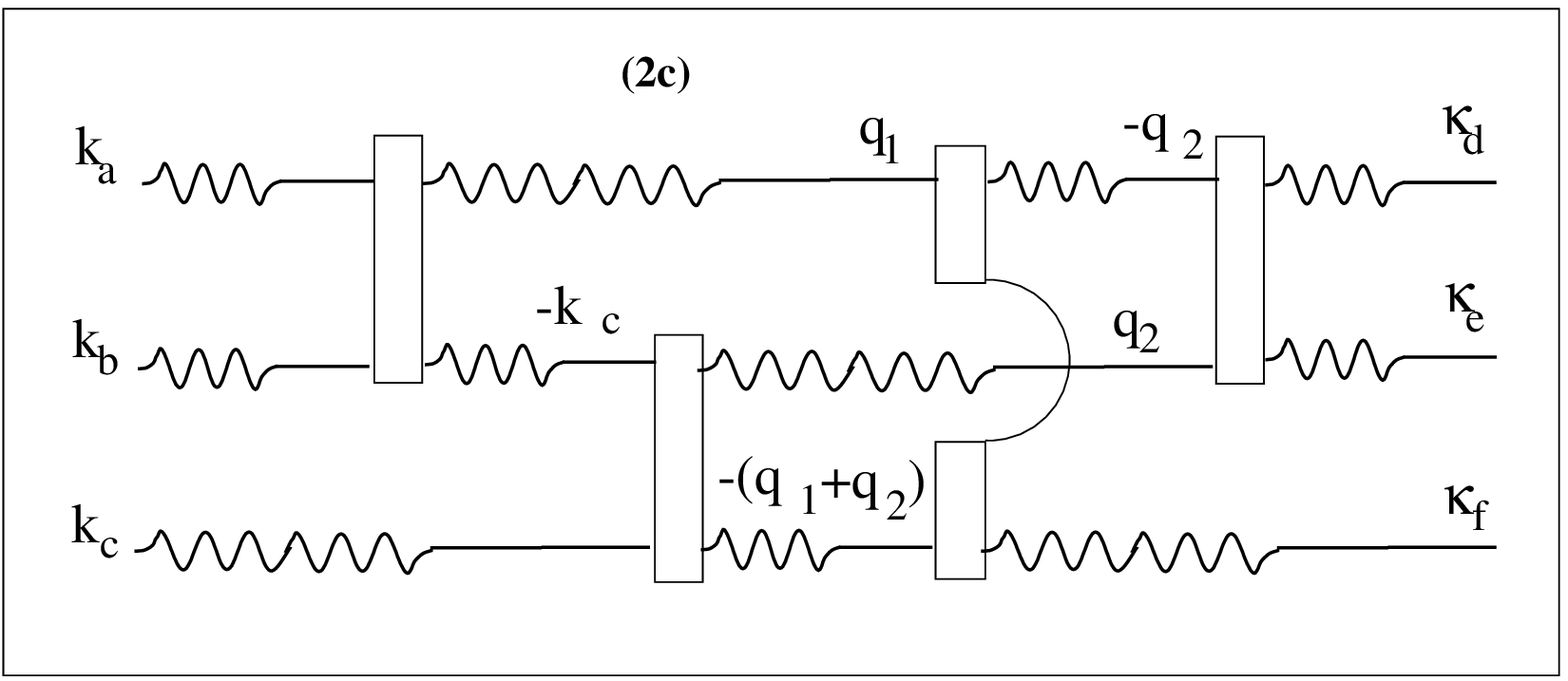}
\epsfxsize=8.6truecm
\epsfbox{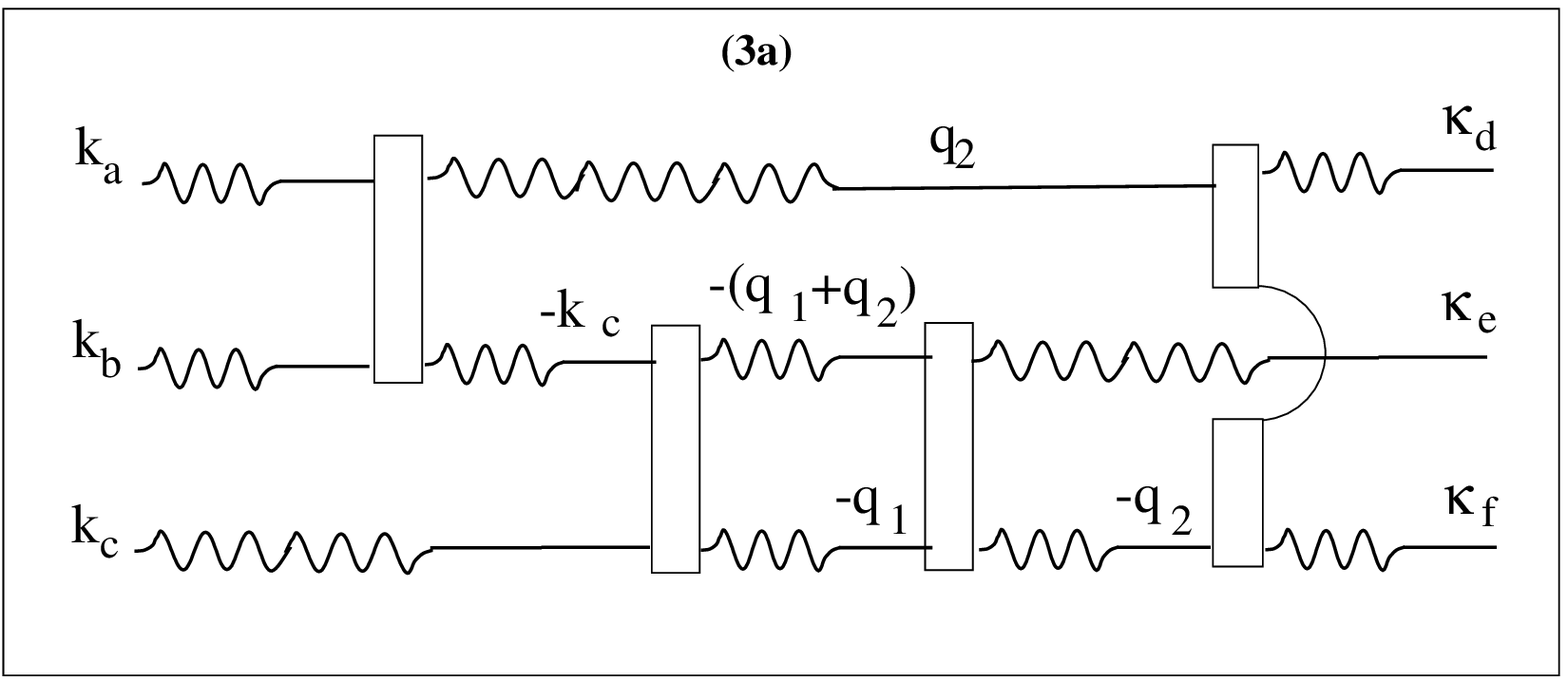}
\epsfxsize=8.6truecm
\epsfbox{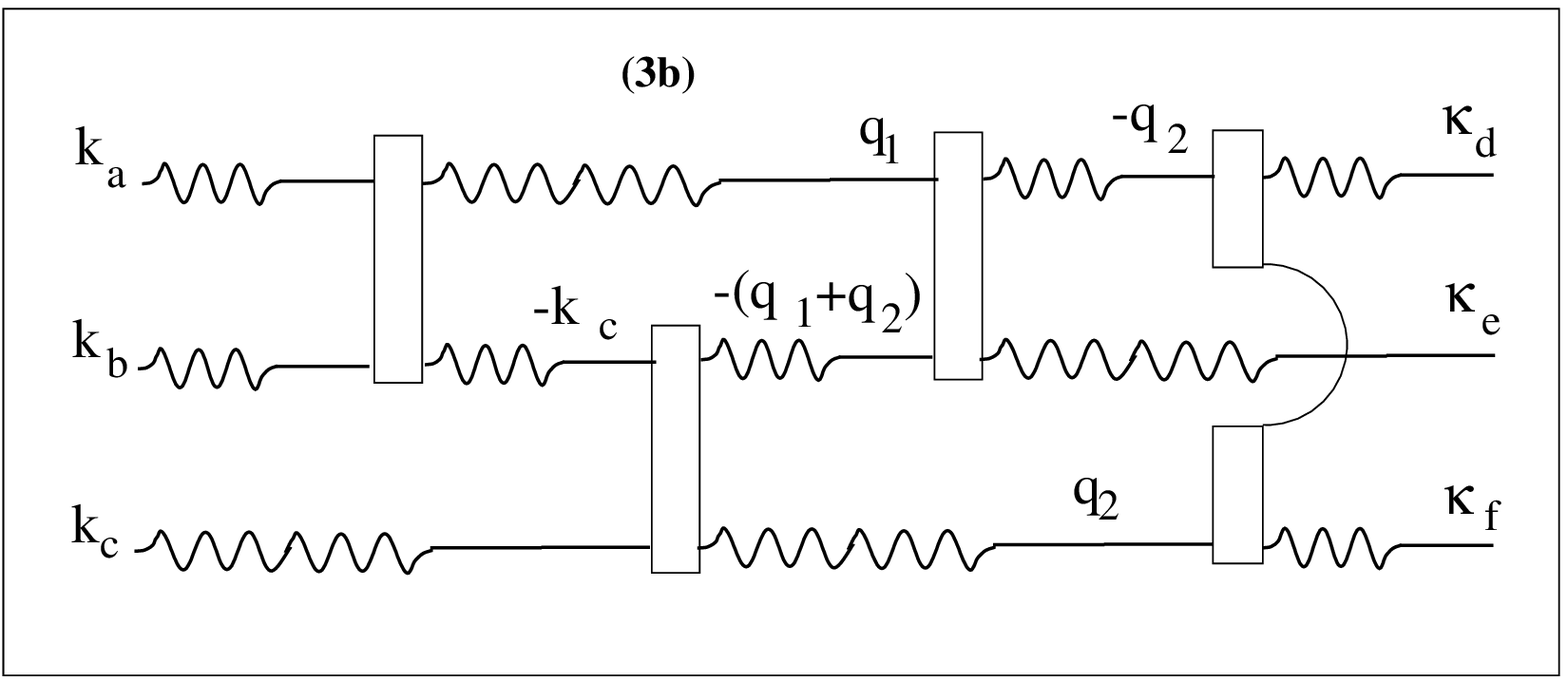}
\epsfxsize=8.6truecm
\epsfbox{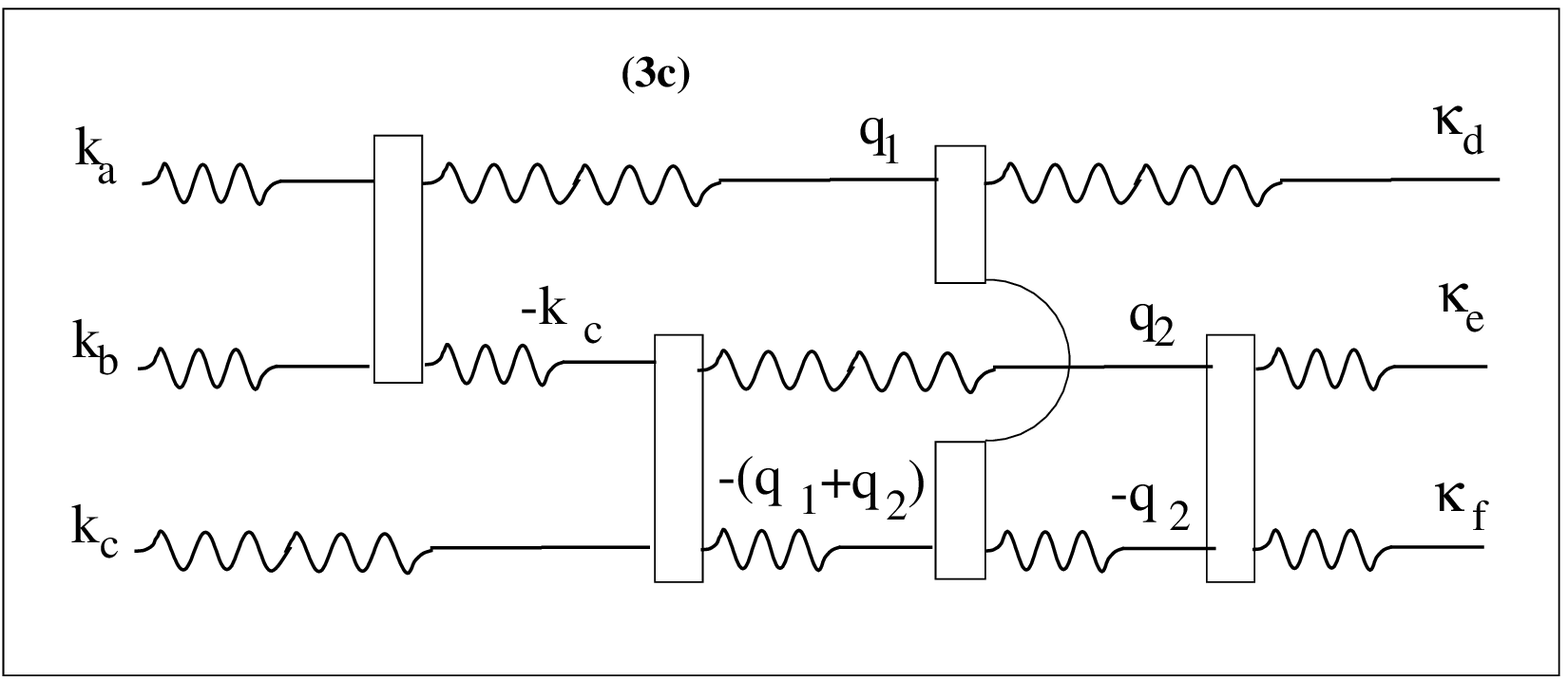}
\vskip 0.3 cm
\caption{The second group of six 2-loop diagrams
 appearing in the loop expansion of
$G_{3,3}$.}
\label{fig-8}
\end{figure}
\begin{figure}
\epsfxsize=8.6truecm
\epsfbox{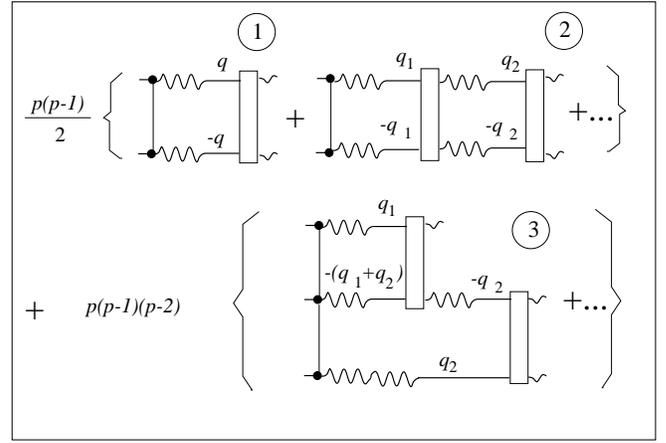}
\vskip 0.3 cm
\caption{The amputated diagrams that appear up to 2-loop in the
loop expansion of $G_{p,p}$, with the appropriate combinatorial
factors.}
\label{fig-9}
\end{figure}

The second group of six diagrams (2a -- 3c) shown in Fig.~\ref{fig-8}
yields to a similar analysis, but the amputated diagram is shown as
diagram (3) in Fig.~\ref{fig-9}. All six diagrams result in the very
same amputation, up to permutations of the three struts. Analyzing the
amputated diagram (3) one brings it to the canonical form
(\ref{2loop}) with $\tilde
\Psi(q_1,q-2)$ given by (\ref{tpsi2}). Accordingly we write
\begin{equation}
T_{3,2a,\dots 3c}^{(2)}(r,\kappa)=6\tilde
\delta^2\left[\frac{1}{2}\ln^2{\Big(\frac{1}
{\kappa r}\Big)}+b_2\ln{\Big(\frac{1}{\kappa
r}\Big)}\right] T_3^{\rm s}(r,\kappa)
\ ,
\label{Kkp2a}
\end{equation}
with $b_2$ of (\ref{cb2}).

The analysis of the 2-loop diagrams that involve 4-point rungs in the
context of $G_{p,p}$ follows exactly the same lines, with the
amputated diagrams being those of Fig.~\ref{fig-9}. The only thing to
mind is the combinatorics, which are presented explicitly in
Fig.~\ref{fig-9}, leading to the numbers in Eq.~(\ref{zpaf}).
\section{Resummed equations for the 4-point and 6-point rungs}
\label{6rung}
In this Appendix we sketch a theory for the 4-point and 6-point rungs.
Our main aim here is to explain why the 6-point rung is quadratic
in the smallness, but we use the opportunity to indicate how a future
theory of these objects may be formulated.

Consider the beginning of the series expansion of the 4-point rung
which is shown in Fig. \ref{fig-1}b. Diagram (2) contains 
a cross of correlators each attached to two 3-point vertices. This
is exactly diagram (1), and therefore the equation can lend itself
to resummation resulting in the equation shown in 
 Fig.~\ref{fig-rungs}a. 
We note that this is not the full equation for the 4-point rung even
in 1-loop order 
\begin{figure}
\epsfxsize=8.6truecm
\epsfbox{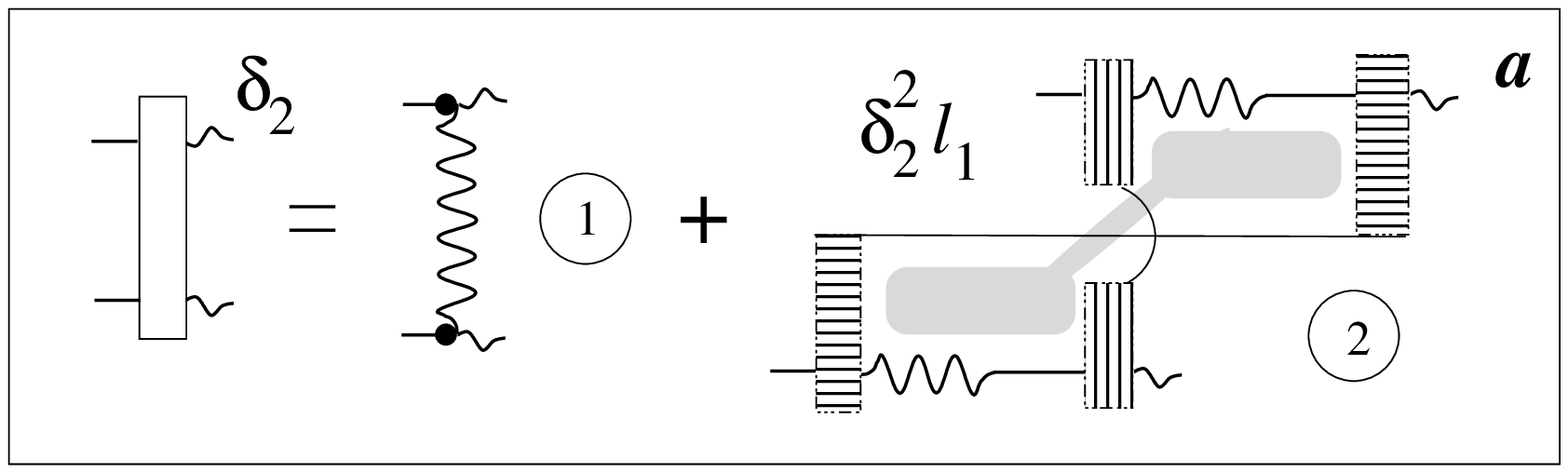}
\epsfxsize=8.6truecm
\epsfbox{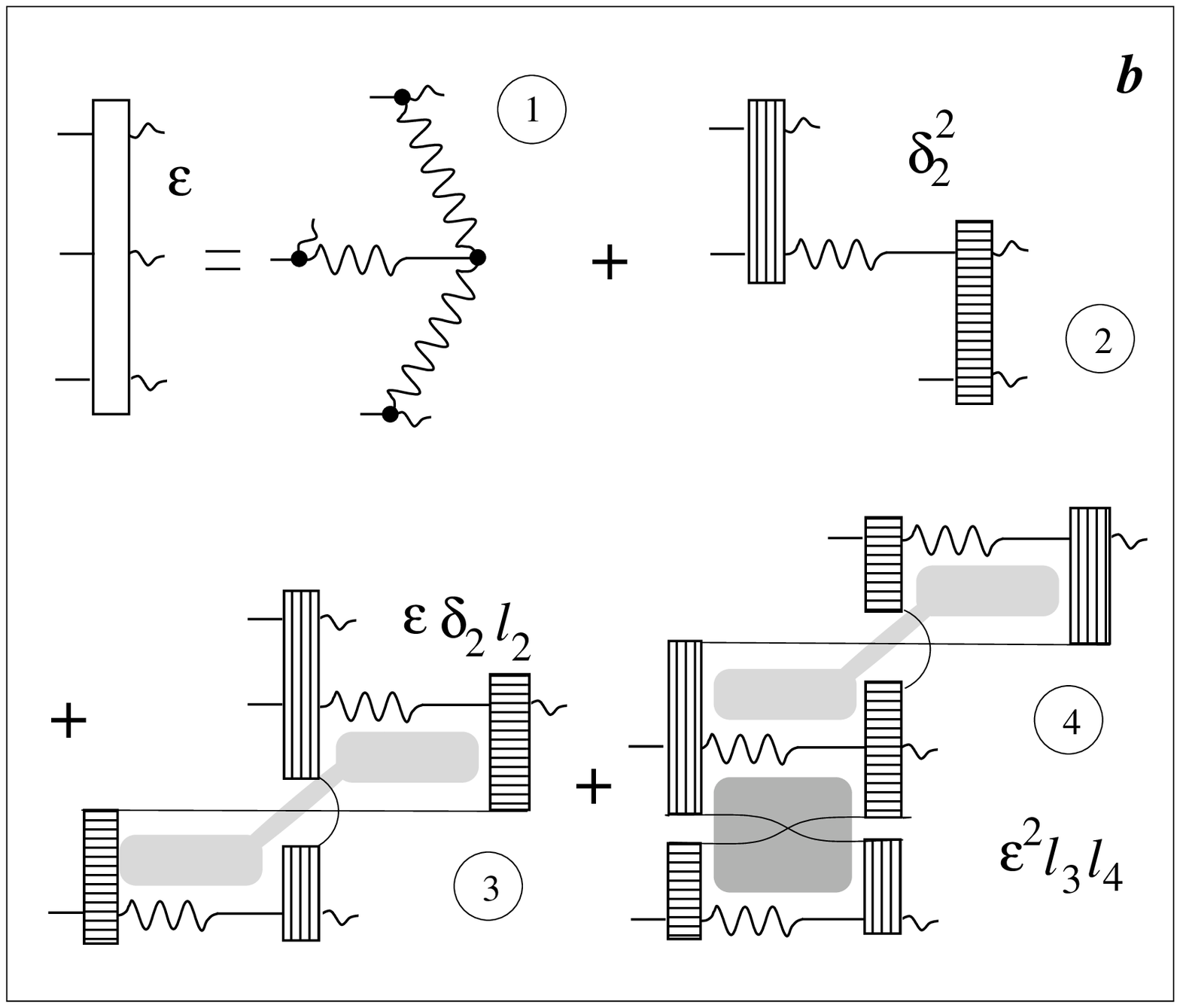}
\vskip 0.3 cm
\caption{A sketch of the equations for the 4-point rung (panel a)
and 6-point rung (panel b). The shading of rungs is used to identify 
different pieces of the same object. In the same manner we shade
the loops to identify one or two loop integrals.}
\label{fig-rungs}
\end{figure}\noindent
since we did not take into account the ladders with a correlator and
Green's function in a cross section. Taking into account all the
needed contributions is not difficult, but is not the main point of
this Appendix, and we proceed for simplicity without the additional
terms.

In the asymptotic regime the bare contribution diagram (1) in
Fig.~\ref{fig-rungs}a is negligible. With this contribution discarded,
the remaining equation is homogeneous, calling for finding a zero mode
of the equation. Since we have already demonstrated that the 4-point
rung is small, of the order of $\delta_2$ we can conclude that the
loop integral which we denote as $\ell_1$ must be large, or the order
of $1/\delta_2$. (the homogeneous equation can be only solved if
$\delta_2\approx \delta_2^2\ell_1$). In fact, in the future it would
be extremely worthwhile to solve the full equation in the 1-loop order
and demonstrate that this is the case, and thus to lend further weight
to the theory presented in this paper. Of course solving such an
equation will also supply us with a functional form of the 4-point
rung, and with it a substantial part of the value of the parameter
$b_2$ which appears in the final result for the scaling exponents.

In Fig.~\ref{fig-rungs}b we present the resummed form of the equation
of the 6-point rung, to the same level of qualitative
discussion. Again we discard in the asymptotic limit the bare
contribution of diagram (1), but we cannot neglect diagram (2) since
it has the same asymptotic behavior as the resummed 6-point
rung. Diagram (2) is of the order of $\delta_2^2$. Diagram (3) is of
the order of $\epsilon\delta_2 \ell_2$ where $\ell_2$ is the loop
integral. This integral is very similar to $\ell_1$, and we therefore
estimate $\ell_2\approx \ell_1 \approx 1/\delta_2$, and thus diagram
(3) is of the order of the LHS. Diagram (4) is of the order of
$\epsilon^2 \ell_3\ell_4$ where $\ell_3$ and $\ell_4$ each refers to
one of the loop integrals. With the same level of approximation we
estimate it thus to be of $O(\epsilon^2/\delta_2^2)$.  Denoting
$x\equiv \epsilon/\delta^2$ we thus represent the order of magnitude
relations that result from panel b by the equation
\begin{equation}
x=1+ax+bx^2 \ ,
\end{equation}
where $a$ and $b$ are dimensionless constants of $O(1)$. It is obvious
that only $x\approx 1$ is a consistent solution of this equation, and
we thus conclude that {\em the 6-point rung is quadratic in the
smallness} $\delta_2$.

We therefore understand that the 6-point rung appears in our
considerations only at the level of the $O(\delta^2)$ order. In this
order it appears in addition to the 2-loops integrals which are formed
by two 4-point rungs, as discussed in detail in the text of the
paper. But since the 6-point rung connects three struts, exactly like
the structure made of {\em two} 4-point rungs, the combinatorical
factors appearing in the $p$th order scaling exponents {\em are
identically the same}. Accordingly we understand that the effect of
the 6-point rung is only in renormalizing the value of the parameter
$b_2$ which anyway is model dependent.

Similar consideration apply to the 8-point rung which begins to affect
the theory only in $O(\delta_2^3)$. It will renormalize the value of
the parameter $b_3$ in Eq.~(\ref{3loopzn}). Higher order rungs are even
less relevant for the calculation at hand.
 
\end{multicols}



 \end{document}